\documentclass[longauth]{aa}  
\usepackage[switch]{lineno}
\usepackage{graphicx}
\usepackage{subcaption}
\usepackage{caption}
\usepackage{float} 
\usepackage{placeins}
\usepackage{tabularx}
\usepackage{array}
\usepackage{CJKutf8}
\usepackage{siunitx}
\usepackage[varg]{txfonts}
\usepackage{afterpage}
\usepackage{CJK}
\usepackage{graphicx}
\usepackage{graphicx}
\usepackage{subcaption}
\usepackage{txfonts}
\usepackage{silence}
\usepackage{ulem}
\usepackage[justification=raggedright,singlelinecheck=false]{caption}
\usepackage{booktabs}
\newcommand{\ha} {\mbox{H$\alpha$}\,}
\newcommand{\hb} {\mbox{H$\beta$}\,}
\newcommand{\hg} {\mbox{H$\gamma$}\,}

\newcommand{\Nai}{\ion{Na}{i}\,}

\newcommand{\Feii} {\ion{Fe}{ii}\,}

\newcommand{\Caii} {\ion{Ca}{ii}\,}

\newcommand{\Cii} {\ion{C}{ii}\,}

\newcommand{\Hei} {\ion{He}{i}\,}
\newcommand{\Heii} {\ion{He}{ii}\,}

\newcommand{\Nii} {\ion{N}{ii}\,}

\newcommand{\Oi} {\ion{O}{i}\,}

\newcommand{\SiII} {\ion{Si}{ii}\,}

\newcommand{\Mgi} {\ion{Mg}{i}\,}

\newcommand{\Scii} {\ion{Sc}{ii}\,}

\newcommand{\Tiii} {\ion{Ti}{ii}\,}
\newcommand{\Baii} {\ion{Ba}{ii}\,}

\usepackage{etoolbox} 
\makeatletter
\patchcmd{\AA@hyperref}
  {\AddToHook{begindocument}{showkeys}{before}{nameref}}
  {}{}{}
\makeatother

\usepackage{hyperref}

\begin{document} 
   \title{SN\,2017ckj: A linearly declining type IIb supernova with a relatively massive hydrogen envelope}
   \titlerunning{SN\,2017ckj}
   \author{L.-H. Li\begin{CJK*}{UTF8}{gbsn}(李陆翰)\end{CJK*}\inst{\ref{inst1},\ref{inst2},\ref{inst3}}
     \and S. Benetti\inst{\ref{inst4}}
     \and Y.-Z. Cai\begin{CJK*}{UTF8}{gbsn}(蔡永志)\end{CJK*}\inst{\ref{inst1},\ref{inst2}}\fnmsep\thanks{Corresponding authors: caiyongzhi@ynao.ac.cn (CYZ)}
     \and B. Wang\begin{CJK*}{UTF8}{gbsn}(王博)\end{CJK*}  \inst{\ref{inst1},\ref{inst2}}\fnmsep\thanks{wangbo@ynao.ac.cn (WB)}
     \and A. Pastorello \inst{\ref{inst4}}
     \and N. Elias-Rosa\inst{\ref{inst4},\ref{inst13}}
     \and A. Reguitti \inst{\ref{inst4},\ref{inst25}}
     \and L. Borsato \inst{\ref{inst4}}
     \and E. Cappellaro \inst{\ref{inst4}}
     \and A. Fiore \inst{\ref{inst4},\ref{inst6}}
     \and M. Fraser \inst{\ref{inst20}}
     \and M. Gromadzki \inst{\ref{inst31}}
     \and J.~Harmanen \inst{\ref{inst19}}
     \and J. Isern \inst{\ref{inst13},\ref{inst27},\ref{inst26}}
     \and T.~Kangas \inst{\ref{inst18},\ref{inst19}}
     \and E. Kankare \inst{\ref{inst19}}
     \and P.~Lundqvist \inst{\ref{inst30}}
     \and S.~Mattila \inst{\ref{inst19},\ref{inst32}}
     \and P. Ochner \inst{\ref{inst4},\ref{inst12}}
     \and Z.-H. Peng\inst{\ref{inst11}}
     \and T.~M.~Reynolds \inst{\ref{inst19},\ref{inst21},\ref{inst22}}
     \and I. Salmaso \inst{\ref{inst14}}
     \and S.~Srivastav\inst{\ref{inst8}}
     \and M.~D.~Stritzinger \inst{\ref{inst17}}
     \and L. Tomasella \inst{\ref{inst4}}
     \and G.~Valerin  \inst{\ref{inst4}} 
     \and Z.-Y. Wang\inst{\ref{inst9},\ref{inst10}}
    \and J.-J. Zhang \inst{\ref{inst1},\ref{inst2}}
     \and C.-Y. Wu\inst{\ref{inst1},\ref{inst2},\ref{inst3}}
    }

\institute{
\label{inst1}Yunnan Observatories, Chinese Academy of Sciences, Kunming 650216, China \and
\label{inst2}International Centre of Supernovae, Yunnan Key Laboratory, Kunming 650216, China \and
\label{inst3}University of Chinese Academy of Sciences, Beijing 100049, China \and
\label{inst4}INAF - Osservatorio Astronomico di Padova, Vicolo dell'Osservatorio 5, 35122 Padova, Italy \and
\label{inst13}Institute of Space Sciences (ICE, CSIC), Campus UAB, Carrer de Can Magrans, s/n, E-08193 Barcelona, Spain \and
\label{inst6}INAF - Osservatorio Astronomico d’Abruzzo, Via Mentore Maggini Snc, 64100 Teramo, Italy \and
\label{inst20}School of Physics, O'Brien Centre for Science North, University College Dublin, Belfield, Dublin 4, Ireland \and
\label{inst31}Astronomical Observatory, University of Warsaw, Al. Ujazdowskie 4, 00-478 Warszawa, Poland \and
\label{inst19}Department of Physics and Astronomy, University of Turku, FI-20014 Turku, Finland \and
\label{inst27}Fabra Observatory, Royal Academy of Sciences and Arts of Barcelona (RACAB), 08001 Barcelona, Spain \and
\label{inst26}Institute for Space Studies of Catalonia (IEEC), Campus UPC, 08860 Castelldefels (Barcelona), Spain \and
\label{inst18}Finnish Centre for Astronomy with ESO (FINCA), University of Turku, Vesilinnantie 5, Quantum 20014 Turku, Finland \and
\label{inst30}The Oskar Klein Centre, Department of Astronomy, Stockholm University, AlbaNova, SE-10691 Stockholm, Sweden \and
\label{inst32}School of Sciences, European University Cyprus, Diogenes Street, Engomi, 1516 Nicosia, Cyprus \and
\label{inst12}Physics and Astronomy Department Galileo Galilei, University of Padova, Vicolo dell’Osservatorio 3, I-35122, Padova, Italy \and
\label{inst11}School of Electronic Science and Engineering, Chongqing University of Posts and Telecommunications, Chongqing 400065, P.R. China \and
\label{inst25}INAF - Osservatorio Astronomico di Brera, Via E. Bianchi 46, 23807 Merate (LC), Italy \and
\label{inst21}Cosmic Dawn Center (DAWN) \and
\label{inst22}Niels Bohr Institute, University of Copenhagen, Jagtvej 128, 2200 København N, Denmark \and
\label{inst14}INAF–Osservatorio Astronomico di Capodimonte, Salita Moiariello 16, 80131 Napoli, Italy \and
\label{inst8}Astrophysics Research Centre, School of Mathematics and Physics, Queen’s University Belfast, Belfast BT7 1NN, UK \and
\label{inst17}Department of Physics and Astronomy, Aarhus University, Ny Munkegade 120, DK-8000 Aarhus C, Denmark \and
\label{inst9}School of Physics and Astronomy, Beijing Normal University, Beijing 100875, P.R. China \and
\label{inst10}Department of Physics, Faculty of Arts and Sciences, Beijing Normal University, Zhuhai 519087, P.R. China
}
   \date{Received September xx, 2025; accepted March xx, 2025}

    \abstract{
    We present optical observations of the type IIb supernova (SN) 2017ckj, covering approximately 180 days after the explosion.
    Its early-time multi-band light curves display no clear evidence of a shock-cooling tail, resembling the behaviour of SN\,2008ax.
    The $V$-band light curve exhibits a short rise time of about 5 days and reaches an absolute fitted peak magnitude of $M_{\rm V}=-18.49\pm0.18\,\mathrm{mag}$. 
    The late-time multi-band light curves reveal a linear decline.    
    We modelled the bolometric light curve of SN\,2017ckj to constrain the progenitor and the explosion parameters.
    We estimated a total mass of $\rm ^{56}Ni$ synthesised by SN\,2017ckj of $M_{\rm Ni} = 0.21^{+0.05}_{-0.03}\ \rm M_\odot$, with a massive H-rich envelope of $M_{\rm env} = 0.4^{+0.1}_{-0.1}\ \rm M_\odot$.
    Both the $\rm ^{56}Ni$ mass and the envelope mass of SN\,2017ckj are higher than those of typical SNe IIb, in agreement with its peculiar light curve evolution. 
    The early-time spectra of SN\,2017ckj are dominated by a blue continuum, accompanied by narrow \ha and \Heii emission lines.
    The earliest spectrum exhibits flash ionisation features, from which we estimated a progenitor mass-loss rate of $\sim 3\times10^{-4}\,\rm M_\odot\, \mathrm{yr}^{-1}$.
    At later epochs, the spectra develop broad P-Cygni profiles and become increasingly similar to those of SNe IIb, especially SN\,2018gk.
    The late-time spectrum at around 139 days does not show a distinct decline in the strength of its \ha emission profile, also indicating a relatively massive envelope of its progenitor.
    Aside from the \ha feature, the nebular spectrum exhibits prominent emission lines of \Oi, \Caii, [\Caii], and \Mgi], which are consistent with the prototypical SN\,1993J.
    }
\maketitle               


\section{Introduction}

Core-collapse (CC) supernovae (SNe) mark the explosive deaths of massive stars with initial masses $\gtrsim 8\,\rm M_\odot $, resulting from the gravitational collapse of their stellar cores \citep[see e.g.][]{2003ApJ...591..288H,2009ARA&A..47...63S}.
A subset of CC SNe originates from progenitors that have lost their outer hydrogen and/or helium envelopes, and these are classified as stripped-envelope supernovae \citep[SE-SNe;][]{1996ApJ...459..547C,2001AJ....121.1648M}.
The mechanism for the envelope stripping is still controversial, with strong stellar winds during the Wolf–Rayet (WR) phase \citep{2008A&ARv..16..209P,2016MNRAS.455..112G} and/or interaction with a binary companion \citep{1992PASP..104..717P,2019NatAs...3..434F} being suggested.
It has been proposed that the stellar winds expected from the single star formation channel would not be able to account for the current observed rate of SE-SNe \citep{2011MNRAS.412.1522S}.
In contrast, some population synthesis studies have indicated that stars evolving in close binary systems can produce a sufficient number of SE-SNe \citep[e.g.][]{2010Yoon,2013Eldri,2019NatAs...3..434F}.

The various subclasses of SE-SNe originate from progenitors that have undergone different degrees of envelope stripping before CC, and exhibit diverse spectral features throughout their spectral evolution \citep[e.g.][]{2017MNRAS.469.2672P}.
If all or the majority of hydrogen is removed from the outer envelope, an H-poor SN Ib/c ultimately occurs.
As a consequence, SNe Ib/c lack any prominent hydrogen features within their spectral evolution and are dominated by helium and metal elements.
However, if the progenitors retain a small amount of hydrogen \citep[$\sim0.001-1.0\,\rm M_\odot$; e.g.][]{2019ApJ...885..130S}, the resulting explosions are classified as SNe IIb.
The mass limit of the hydrogen envelope in SN IIb progenitors remains under debate \citep[e.g.][]{2012MNRAS.422...70H,2017ApJ...840...10Y,2022MNRAS.511..691G}.

The defining characteristic of SNe IIb is the appearance of strong \Hei P-Cygni lines following an early optical spectrum dominated by hydrogen \citep{2017suex.book.....B}.
Due to the presence of a small amount of hydrogen in the envelope of the progenitor, SNe IIb exhibit hydrogen features in their early-time spectra.
These hydrogen features gradually fade over time, and the nebular spectra are similar to those of SN Ib/c, with strong emission lines of \Mgi], [\Oi] and [\Caii].
SN\,1987K was the first SN observed to undergo a spectral transition from type II to type Ib, and was subsequently classified as a SN IIb \citep{1988AJ.....96.1941F}.
Subsequently, SN\,1993J was discovered on 28.9 March 1993 in the nearby galaxy M81 \citep[][]{1993IAUC.5731....1R}.
SN\,1993J has been extensively studied and is considered a prototypical example of SNe IIb \citep[e.g.][]{1994ApJ...429..300W,2003astro.ph.10228F}.

The double-peaked light curves, primarily observed in the optical bands, have been reported in a few SNe IIb, such as SN\,1993J \citep[e.g.][]{1994AJ....107.1022R}, SN\,2011dh \citep[e.g.][]{2013MNRAS.433....2S,2014A&A...562A..17E,2014ApJ...781...69M,2015A&A...580A.142E}, SN\,2011fu \citep{2015MNRAS.454...95M}, SN\,2016gkg \citep{2018Natur.554..497B}, SN\,2024uwq \citep{2025arXiv250502908S}, and SN\,2024aecx \citep{2025arXiv250519831Z}.
These SNe IIb are characterised by an initial decline due to shock cooling, followed by a second peak powered by the radioactive decay of $\rm ^{56}Ni$, which typically dominates late-time luminosity evolution.
The shock-cooling emission peak has a mean rise time of $2.07\pm1.0$ days in the $g$ band and lasts only about one-third of the duration of the rise to the $\rm ^{56}Ni$-powered peak \citep{2025ApJ...989..192C,2025A&A...701A.128A}.
However, some SNe IIb with continuous early-time coverage, such as SN\,2008ax \citep[e.g.][]{2008MNRAS.389..955P,2011MNRAS.413.2140T} and SN\,2020acat \citep{2022MNRAS.513.5540M}, do not display an initial shock-cooling tail prior to a $\rm ^{56}Ni$-powered peak.
This observational diversity suggests the existence of two distinct categories of SN IIb progenitors: extended and compact \citep[e.g.][]{2010ApJ...711L..40C,2017suex.book.....B}.
Extended progenitors with large H-rich envelopes exhibit a noticeable shock-cooling tail after the explosion, whereas compact progenitors show no such feature.

Supernova progenitor constraints have been derived through deep pre- and post-explosion imaging with high spatial resolution, primarily using data obtained by the Hubble Space Telescope (HST) and other ground- or space-based facilities.
To date, some SN IIb progenitors have been identified: SN\,1993J \citep[e.g.][]{1994AJ....107..662A,2004Natur.427..129M}, SN\,2008ax \citep{2008MNRAS.391L...5C,2015ApJ...811..147F}, SN\,2011dh \citep{2011ApJ...741L..28V}, SN\,2013df \citep{2014AJ....147...37V}, SN\,2016gkg \citep{2017MNRAS.465.4650K,2017ApJ...836L..12T}, SN\,2017gkk \citep{2024ApJ...970L...9N}, and SN\,2024abfo \citep{2025A&A...698A.129R,2025ApJ...987L..10N}.
The progenitor of SN\,1993J was identified in pre-explosion imaging as a K-type supergiant \citep{2002PASP..114.1322V}.
A B-type companion star was later detected, providing direct evidence that SN\,1993J originated from a binary system undergoing a mass-transfer phase \citep{2004Natur.427..129M}.
Yellow supergiant (YSG) progenitors with an initial mass of $\sim 10-17\,\rm M_\odot$ have been identified for SN\,2011dh, SN\,2013df, SN\,2016gkg, SN\,2017gkk, and SN\,2024abfo \citep[e.g.][]{2011ApJ...739L..37M,2012ApJ...757...31B,2015ApJ...807...35M,2022ApJ...936..111K,2024ApJ...970L...9N,2025A&A...698A.129R}.
In contrast, the progenitor of SN\,2008ax is likely a highly stripped star (a low-mass analogue of a WR star) in a binary system \citep{2008MNRAS.391L...5C,2015ApJ...811..147F}.
Another SN IIb, SN\,2013cu, was suggested to have a WR–like progenitor undergoing intense mass loss shortly before the explosion, based on the detection of narrow high-ionisation emission lines in its early flash spectrum \citep{2014Natur.509..471G}.

In this paper, we present photometric and spectroscopic observations of SN\,2017ckj, a peculiar and luminous SN IIb characterised by a linear decline in its multi-band light curves.
In Section~\ref{sect:basic_inf}, we report the distance, extinction, and reddening associated with the host galaxy of SN\,2017ckj. 
The photometric and spectroscopic analyses are presented in Sections~\ref{sect:photometry} and~\ref{sect:spectroscopy}, respectively. 
Finally, the discussions are presented in Section~\ref{sect:discussion}, and the concluding remarks are provided in Section~\ref{sect:remarks}.
Additionally, we present the supplementary figure in Appendix~\ref{sect:supfigure}, the data reduction techniques in Appendix~\ref{sect:data reduction}, and the relevant data tables in Appendix~\ref{Sect:datainfo}.

\section{Basic target information} \label{sect:basic_inf}

\begin{figure}
   \centering
   \includegraphics[width = \columnwidth]{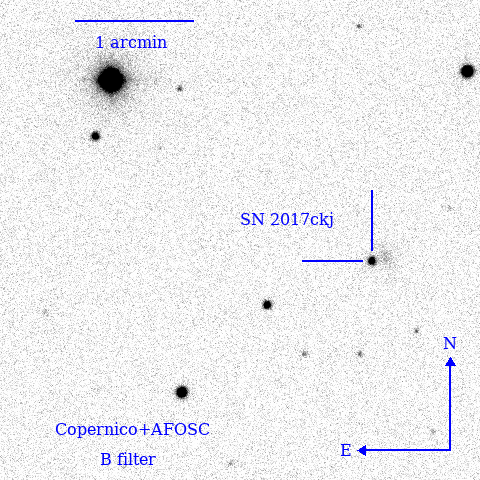}
      \caption{Image of the location of SN\,2017ckj, taken on 29 March 2017 by the Copernico telescope with the $B$ filter. 
      The orientation and scale of the images are reported.}
     \label{Fig:location}
\end{figure}

\begin{table}[htbp]
\centering
\caption{Summary of the basic properties of SN\,2017ckj}
\label{tab:17ckj_properties}
\renewcommand{\arraystretch}{1.1}
\begin{tabular}{@{}lr@{}} 
\hline
RA (J2000) & 16:09:44.40\\
DEC (J2000) & +48:02:07.26\\
Host Galaxy & WISEA\ J160943.68+480209.2\\
Redshift $z$ & $0.037077\pm0.000014$\\
$\mu$ & $35.99 \pm 0.15 \, \mathrm{mag}$\\
$d$ & $158.1 \pm 11.1 \, \mathrm{Mpc}$\\
$E(B - V)_{\mathrm{Gal}}$$^{\mathrm{1}}$ & $0.013 \, \mathrm{mag}$\\
$E(B - V)_{\mathrm{host}}$ & $\sim0 \, \mathrm{mag}$\\
Last Non-Detection (MJD) & 57834.62\\
First Discovery (MJD) & 57838.55\\
Midpoint epoch (MJD) & $57836.6\, \pm 2.0$\\
Estimated Explosion (MJD)$^{\mathrm{2}}$ & $57837.1\, \pm 0.1$\\
\hline
\end{tabular}
\vspace{0.1mm}
\begin{flushleft}
Notes:
1. Retrieved from NASA/IPAC NED \citep{Schlafly2011ApJ...737..103S}. \\
2. Estimated from the shock-cooling light curve fitting.
\end{flushleft}
\end{table}

SN\,2017ckj was discovered by the Asteroid Terrestrial-impact Last Alert System \citep[ATLAS;][]{Tonry2018ApJ...867..105T,Tonry2018PASP..130f4505T, Smith2020PASP..132h5002S} on 26.55 March 2017 (epoch corresponding to MJD = 57838.55; UT dates are used throughout this paper), at an ATLAS cyan-filter ($c$) brightness $c = 17.81\, \mathrm{mag}$ \citep{2017TNSTR.355....1T}. 
The last non-detection by ATLAS was on 22.62 March 2017 (MJD = 53834.62) in $c$ band, with an estimated limit of 19.53\,mag.
Soon after its discovery, it was classified as a CC SN event by \citet{2017ATel10219....1T} and \citet{2017TNSCR.373....1B} in the framework of the Asiago Transient Classification Program \citep{2014AN....335..841T}.
Its J2000 coordinates are $\rm \alpha = 16^{h}09^{m}44.400^{s}$, $\rm \delta = + 48^{\circ}02^{\prime}07.26^{\prime\prime}$, placing it $2.01^{\prime\prime}$ south and $10.70^{\prime\prime}$ east of the core of the host galaxy WISEA J160943.68+480209.2 
\citep[also named SDSS J160943.66+480209.4;][]{2013wise.rept....1C}.
The location of SN\,2017ckj within the host galaxy is illustrated in Figure~\ref{Fig:location}, and the basic properties of SN\,2017ckj are shown in Table~\ref{tab:17ckj_properties}.

The redshift of the host galaxy of SN\,2017ckj is $0.037077 \pm 0.000014$.
Adopting a standard cosmology \citep[$\rm H_0=73\pm5\,kms^{-1} \, Mpc^{-1}$, $\Omega_M=0.27$ and $\Omega_{\Lambda}=0.73$;][]{Spergel2007ApJS..170..377S} and corrected for the Virgo Cluster, the Great Attractor, and the Shapley supercluster influence, we obtained a Hubble flow distance $d = 158.1\pm11.1\, \mathrm{Mpc}$ and distance modulus $\mu = 35.99\pm0.15\,\mathrm{mag}$ for WISEA J160943.68+480209.2 \footnote{\url{https://ned.ipac.caltech.edu}} \citep{2017ApJS..233...25A}.
Regarding the interstellar reddening, we adopt $E(B-V)_{\rm Gal} = 0.013$\,mag for the Galactic reddening contribution \citep{Schlafly2011ApJ...737..103S}, retrieved via the NASA/IPAC Extragalactic Database (NED), assuming a reddening law with $R_V = 3.1$ \citep{Cardelli1989ApJ...345..245C}.
No strong, narrow interstellar \Nai D $\lambda\lambda$5890,5896 absorption lines were detected at the redshift of the host galaxy.
The lack of strong \Nai D lines, along with the position of SN\,2017ckj relative to its host galaxy, implies that the dust extinction of the host galaxy $E(B-V)_{\rm host}$ is negligible.
Therefore, we adopt $E(B-V)_{\rm total} = 0.013$\,mag as the total reddening towards SN\,2017ckj.

\section{Photometry} \label{sect:photometry}

\subsection{Apparent light curves} \label{Sect:Apperent_LC}
\begin{figure} 
   \centering
   \includegraphics[width = 1\columnwidth]{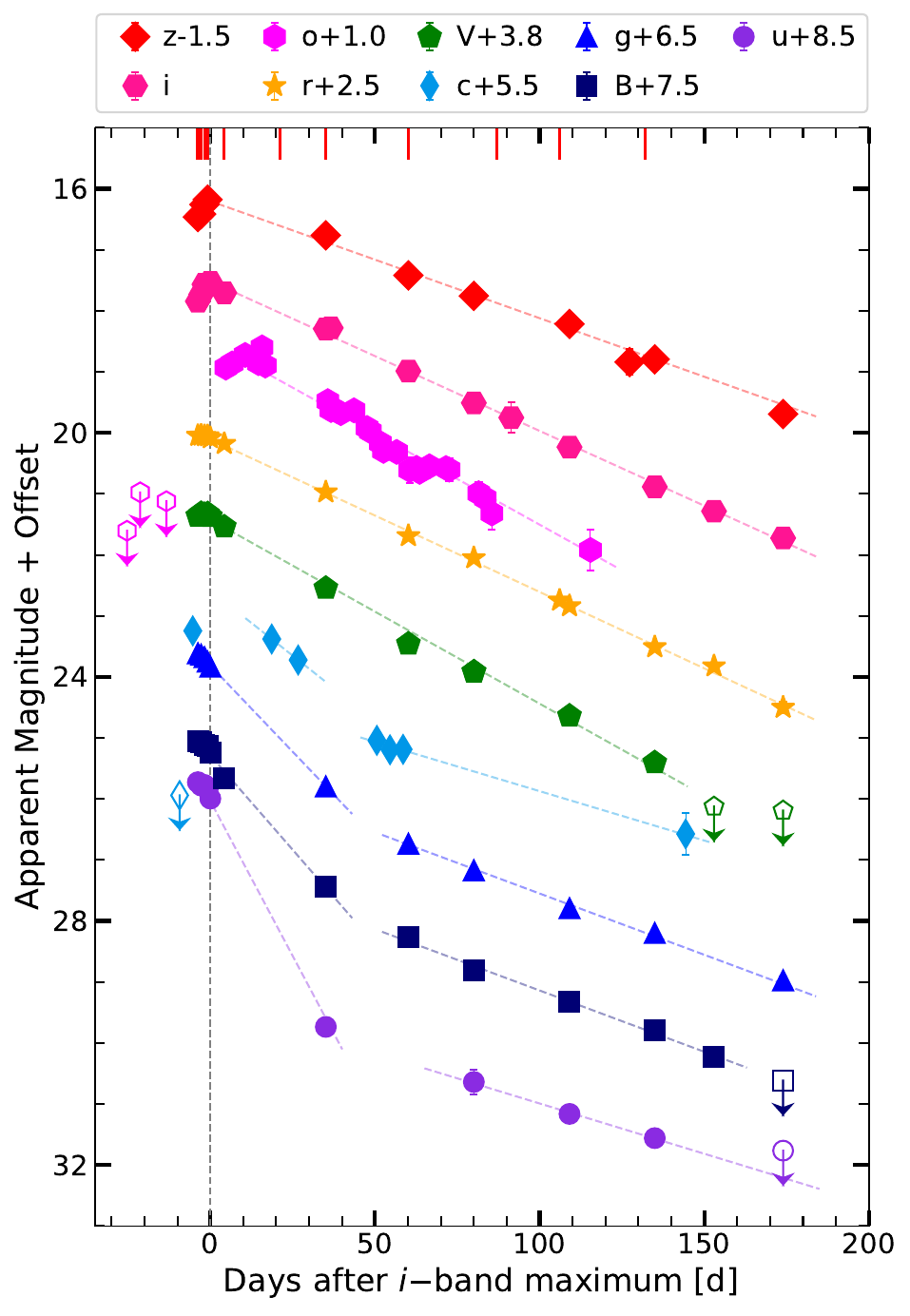}
   \caption{Multi-band light curves of SN\,2017ckj. 
   A dashed vertical line is used to visually represent the reference epoch (MJD = 57843.9), which corresponds to the $i$-band observation maximum. 
   The linear fit is applied to the light curve in each band, as indicated by the dashed lines.
   The epochs of our spectra are marked with vertical solid red lines on the top.
   In most cases, the errors associated with the magnitudes are smaller than the plotted symbol sizes.}
    \label{Fig:LC}
\end{figure}

\begin{table}
    \centering
    \caption{Decline rates (in $\rm mag\, /100\,d$) of the multi-band light curves of SN\,2017ckj.}
    \renewcommand{\arraystretch}{1.2}
    \setlength{\tabcolsep}{15pt}
    \begin{tabular}{ccc}
        \hline
        Filter & $\gamma_{0-40}$ & $\gamma_{40-200}$ \\
        \hline
        u & 10.16$\pm$0.27 & 1.65$\pm$0.40 \\
        B & 6.19$\pm$0.11 & 2.01$\pm$0.11 \\
        g & 5.65$\pm$0.08 & 2.01$\pm$0.04 \\
        c & 4.28$\pm$0.59 & 1.62$\pm$0.38 \\
        \hline
        Filter & \multicolumn{2}{c}{$\gamma_{0-200}$} \\
        \hline
        V & \multicolumn{2}{c}{3.02$\pm$0.02} \\
        r & \multicolumn{2}{c}{2.50$\pm$0.01} \\
        o & \multicolumn{2}{c}{2.99$\pm$0.09} \\
        i & \multicolumn{2}{c}{2.47$\pm$0.02} \\
        z & \multicolumn{2}{c}{1.92$\pm$0.03} \\
        \hline
    \end{tabular}
    \label{tab:decline_rate}
\end{table}

\begin{figure*}
    \centering
    \includegraphics[width=0.95\textwidth]{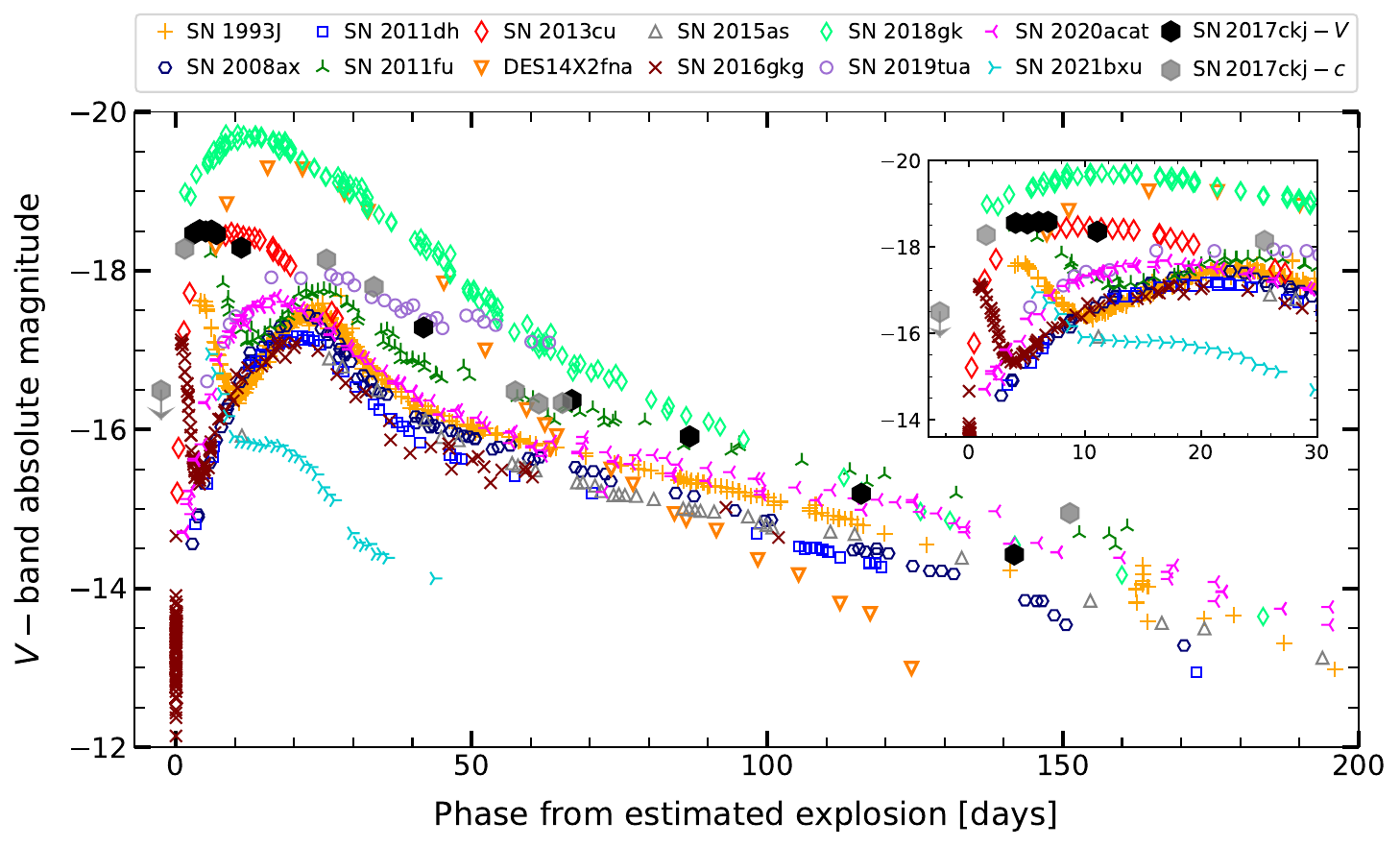}
    \caption{Absolute $V$-band light curve of SN\,2017ckj compared to other SNe IIb.
            The ATLAS absolute $c$-band data of SN\,2017ckj are also plotted with grey prismatic dots, as the $V$-band data is missing from 15 to 40 days.
            The subplot (upper right) displays the initial 30 days of absolute light curves.
            Note that DES14X2fna and SN\,2013cu lack the $V$-band observations, therefore, we substitute with $r$-band data. }
    \label{fig:Abs_V}
\end{figure*}

\begin{figure*}
   \centering
   \includegraphics[width = 1\textwidth]{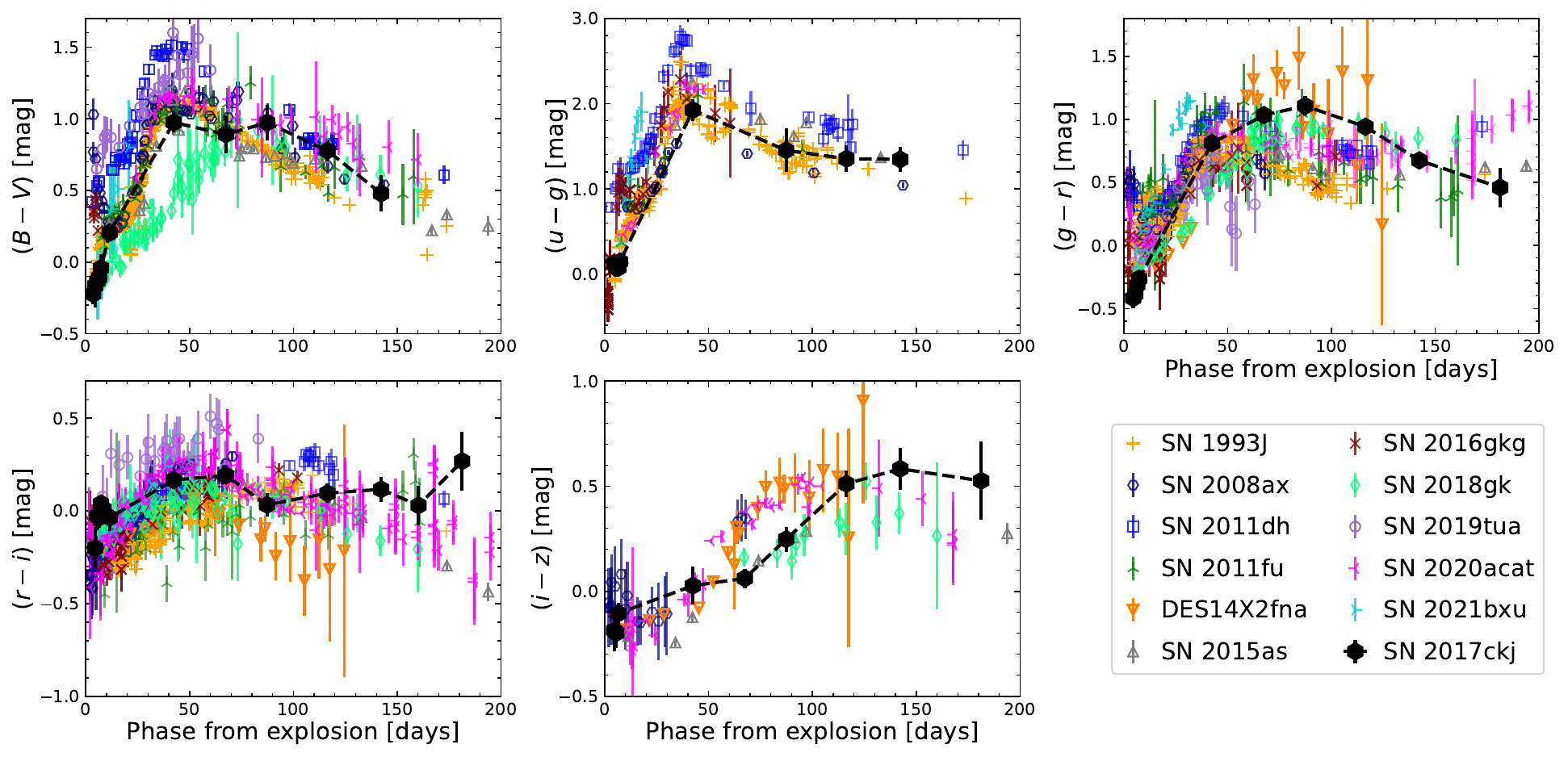}
   \caption{Colour evolution of SN\,2017ckj based on the estimated explosion epoch, compared with those of a sample of SNe IIb. 
   The colour curves are corrected for Galactic and host galaxy extinction. 
   }
    \label{Fig:color_evo}
\end{figure*} 

We conducted continuous photometric observations of SN\,2017ckj for about six months after its discovery.
The multi-band optical light curves of SN\,2017ckj are shown in Figure~\ref{Fig:LC}.
For SN\,2017ckj, the last non-detection $t_l$ occurred on MJD~=~57834.62 with an estimated limit of 19.53$\rm\,mag$ in the ATLAS $c$-band, while the first detection epoch $t_d$ is MJD~=~57838.55 with $c = 17.81\, \mathrm{mag}$ \citep{2017TNSTR.355....1T}. 
The midpoint between $t_l$ and $t_d$ provides a rough estimate of the explosion epoch of MJD\,=\, 57836.6$\,\pm\,$2.0.
To more accurately determine the explosion epoch, we fitted the early-time light curve using the Light Curve Fitting package
\footnote{\url{https://github.com/griffin-h/lightcurve_fitting}}
\citep{hosseinzadeh_2024_11405219}, which implements the shock-cooling model of \citet{2017ApJ...838..130S}.
This package has been applied to shock-cooling analyses of several SNe, including SN\,2016gkv \citep{2018ApJ...861...63H}, SN\,2021yja \citep{2022ApJ...935...31H}, and SN\,2023ixf \citep{2023ApJ...953L..16H}.
We ran 100 walkers for 5000 steps until convergence, which was confirmed through visual inspection of the MCMC chains.
The best-fit light curve, along with the posterior distributions and parameter correlations, is shown in Figure~\ref{Fig:shockcooling_fitting.pdf} of Appendix~\ref{sect:supfigure}.
The fitting yields a more precise explosion epoch of MJD\,=\,57837.1$\,\pm\,$0.1, which is adopted in the following analysis.


In Figure~\ref{Fig:LC}, the $B$, $V$ and $i$ bands exhibit a visible peak in their light curves.
To estimate the peak magnitude of SN\,2017ckj, we performed a Markov Chain Monte Carlo (MCMC) fit using a second-order polynomial to model the early-time $BVi$-band light curves.
The peak of the $B$ band is $\rm17.36\pm0.02\,mag$ at $\rm MJD=57840.20\pm 1.6$.
The peak of the $V$ band is $\rm17.54\pm0.03\,mag$ at $\rm MJD=57842.1\pm 1.5$.
The peak of the $i$ band is $\rm17.52\pm0.02\,mag$ at $\rm MJD=57844.7\pm 0.2$.
Note that the ATLAS $o$-band data exhibits a slight upward trend around 5-10\,d after the $i$-band maximum, likely attributable to the increasing contribution from the emission around \ha feature, which lies within the spectral coverage of the $o$ band.

Using the previously fitted peak time, the $V$-band rise time of SN\,2017ckj is approximately 5\,d, which is faster than all SNe IIb.
\citet{2019MNRAS.488.4239P} analysed the light curve morphology of 22 SNe IIb and reported a weighted average rise time of $19.0\pm1.8\,$d. 
SN\,2004ff, a type IIb SN with a rapid $B$-band rise time of $9.4\pm4.0\,$d, made it one of the fastest-rising SNe IIb \citep{2018A&A...609A.134S}.
Similarly, SN\,2013cu had a fast $r$-band rise time of $\sim10\,$d and has been suggested to originate from a WR-like progenitor \citep{2014Natur.509..471G}.
The rapid rise time of SN\,2017ckj indicates that it has originated from a progenitor with unusual characteristics.

In Figure~\ref{Fig:LC}, after reaching peak brightness, the multi-band light curves of SN\,2017ckj exhibit a clear linear decline, particularly in $riz$ bands.
However, the decline rates show noticeable variations around 40 days, especially in the $uBg$ bands.
We therefore estimated the post-maximum decline rates of SN\,2017ckj in different bands by performing a linear regression on the post-peak data, as summarised in Table~\ref{tab:decline_rate}. 
Due to the noticeable change in the light-curve slope around 40\,d in the $uBgc$ bands, we calculated the decline rates from the peak to 40\,d ($\gamma_{0-40}$) and from 40\,d to 200\,d ($\gamma_{40-200}$) for these bands and the overall decline rates for $Vroiz$ bands.
The bluer light curves decay more rapidly than the redder ones during the early phases.
For instance, in the early-time light curve, the $u$ band shows a high decline rate of $\rm \sim10\,mag/100\,d$, while the $z$ band exhibits a much slower rate of only $\rm \sim1.9\,mag /100\,d$.
The late-time decline rates across different bands are relatively shallow, with values around $2\,\mathrm{mag/100\,d}$.

The $r$-band decline rate $\gamma_{0-200}$ of SN\,2017ckj is $\rm2.5\,mag/100\,d$, which approaches the upper limit of the distribution seen in some SNe IIb samples \citep[$\sim\rm1.3-2.5\,mag/100\,d$;][]{2021MNRAS.505.3950G}.
Notably, DES14X2fna and SN 2018gk, both luminous SNe IIb, exhibit high $r$-band decline rates of $4.30\pm0.10$ and $3.03\pm0.06$, respectively, which are significantly higher than those of typical SNe IIb \citep{2021MNRAS.505.3950G}.
\citet{2019MNRAS.488.4239P} calculated the magnitude difference between 40 and 30\,d post-explosion ($\Delta m_{40-30}$) and proposed that this value for SNe IIb in $B$ band ranges from $\sim$0.6 to 1.2.
For SN\,2017ckj, $\Delta m_{40-30}$ of $B$ band is $\sim0.6$ through extrapolation from the decline rate $\gamma_{0-40}$, which lies near the lower end of this range.

\subsection{Absolute light curves}  

Taking into account the distance and reddening estimates reported in Section~\ref{sect:basic_inf}, we calculated the $V$-band peak absolute magnitude of SN\,2017ckj as $M_{\mathrm{V}}=-18.51\pm0.23\,\mathrm{mag}$ based on the direct photometric $V$-band data.
Adopting the fitted peak magnitude from Section~\ref{Sect:Apperent_LC} yields a fitted peak absolute magnitude of $ -18.49 \pm 0.18$~mag.
The absolute $V$-band magnitude light curve of SN\,2017ckj was compared to those of 
SN\,1993J \citep{1994AJ....107.1022R,1995A&AS..110..513B,1996AJ....112..732R}, 
SN\,2008ax \citep{2008MNRAS.389..955P,2009PZ.....29....2T,2011MNRAS.413.2140T}, 
SN\,2011dh \citep{tsvetkov2012photometric,2013MNRAS.433....2S,2014Ap&SS.354...89B}, 
SN\,2011fu \citep{2013MNRAS.431..308K},
SN\,2013cu \citep{2014Natur.509..471G},
DES14X2fna \citep{2021MNRAS.505.3950G},
SN\,2015as \citep{2018MNRAS.476.3611G}, 
SN\,2016gkg \citep{2017ApJ...837L...2A,2018Natur.554..497B},
SN\,2018gk \citep{2021MNRAS.503.3472B},
SN\,2019tua \citep{2024ApJ...970..103H}, 
SN\,2020acat \citep{2022MNRAS.513.5540M},
SN\,2021bxu \citep{2023MNRAS.524..767D}.
These SNe IIb were chosen as comparison objects for SN\,2017ckj as they all possess comprehensive photometric and spectroscopic data around peak time.
The basic properties and information of those SNe IIb are listed in Table~\ref{tab:SNe_IIb info} of Appendix~\ref{Sect:datainfo}.

In Figure~\ref{fig:Abs_V}, we presented the comparison of the absolute $V$-band magnitudes for a subset of these SNe IIb samples.
Note that we additionally plotted the absolute $c$-band data (grey hexagonal dots) of SN\,2017ckj to supplement the observational coverage.
The $V$-band peak absolute magnitude of SN\,2017ckj ($M_{\mathrm{V}}=-18.58\pm0.17\,\mathrm{mag}$) is brighter than most typical SNe IIb, such as SN\,1993J \citep[$M_{\mathrm{V}}=-17.57\pm 0.17$\,mag;][]{1994AJ....107.1022R}, SN\,2008ax \citep[$M_{\mathrm{V}} =-17.61\pm 0.43$\,mag;][]{2011MNRAS.413.2140T}, SN\,2011fu \citep[$M_{\mathrm{V}}=-17.76\pm0.15$\,mag;][]{2015MNRAS.454...95M}, SN\,2015as \citep[$M_{\mathrm{V}} =-16.82\pm0.18$\,mag;][]{2018MNRAS.476.3611G}, and SN\,2020acat \citep[$M_{\mathrm{V}} =17.62\pm0.11$\,mag;][]{2022MNRAS.513.5540M}.
However, SN\,2017ckj is fainter than both DES14X2fna \citep[$M_{\mathrm{r}}= -19.37\pm0.05$\,mag;][]{2021MNRAS.505.3950G} and SN\,2018gk \citep[$M_{\mathrm{V}}= -19.70\pm0.27$\,mag;][]{2021MNRAS.503.3472B}.

In the subplot of Figure~\ref{fig:Abs_V}, some typical SNe IIb, such as SN\,1993J, SN\,2011fu, and SN\,2016gkg, exhibit light curve evolution characterised by an initial decline followed by a rise to a peak around 20\,d after the estimated explosion epoch.
The initial decline observed in these SNe IIb is caused by the shock-cooling phase of the thin H-rich envelope, while the subsequent rise phase is powered by radioactive heating from $\rm ^{56}Ni$ decay \citep{2014ApJ...788..193N}.
However, not all SNe IIb exhibit evidence of a shock-cooling tail, likely due to their lower envelope masses or more compact progenitor (lower radius).
For SN\,2017ckj, the light curve peaks earlier and reaches a relatively higher luminosity, distinguishing it from most typical SNe IIb.
It is worth noting that the brightness of SN\,2017ckj compared to SN\,2013cu, DES14X2fna, and SN\,2018gk is relatively high for typical SNe IIb, and the trends in their light curves are remarkably similar, particularly for SN\,2013cu.

\subsection{Colour evolution}

\begin{figure*}
   \centering
   \includegraphics[width = 0.85\textwidth]{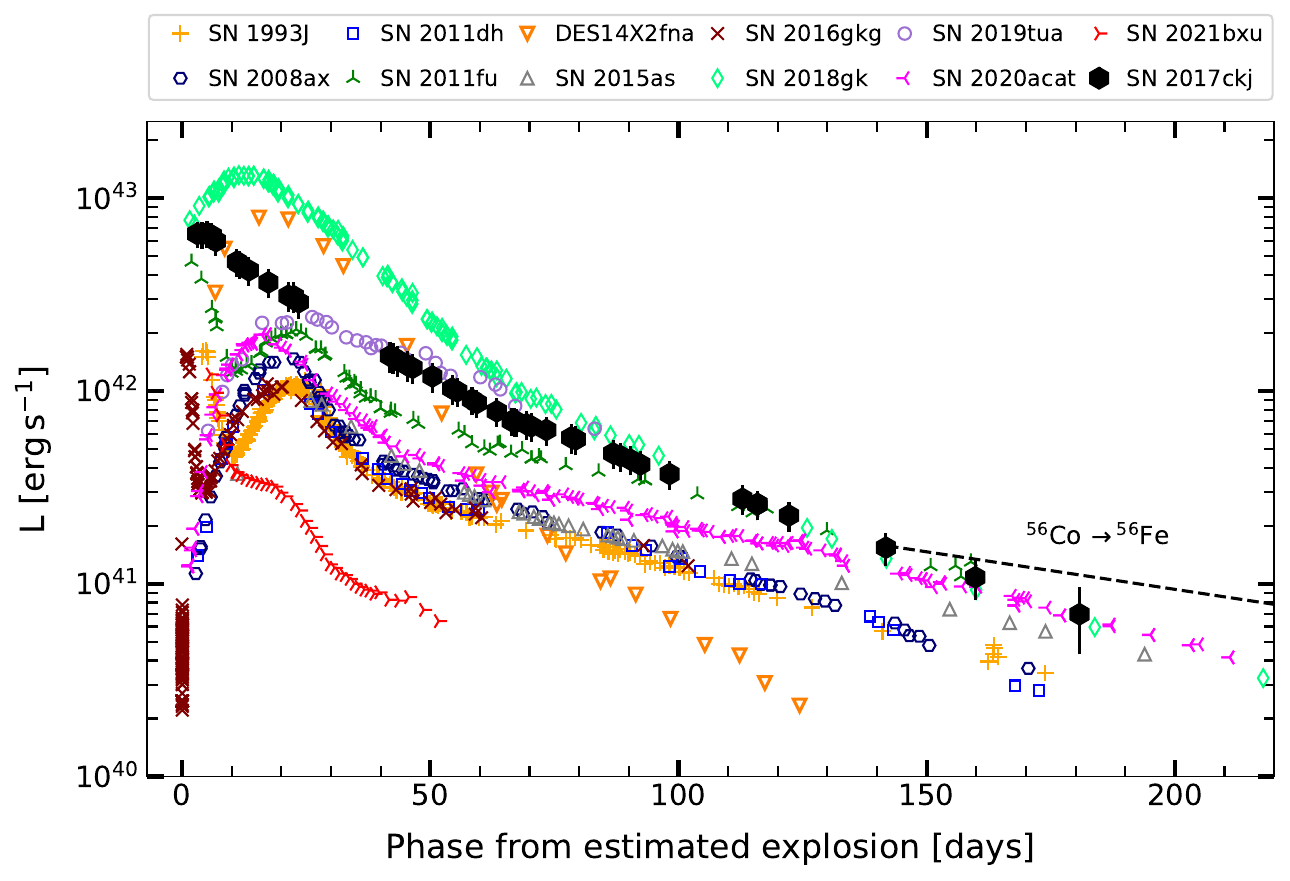}
   \caption{Pseudo-bolometric light curve of SN\,2017ckj, along with several other well-observed SNe IIb over the first 200\,d.
   All SNe have been corrected for reddening as well as time dilation.
   }
    \label{Fig:BL}
\end{figure*}  

For the colour evolution, due to the differences in the observed photometric bands among the SNe IIb samples, additional data were added by using the colour conversion from \citet{2006A&A...460..339J}, of 
\begin{align}
\label{equ:color transform1}
(u-g) &= (0.750 \pm 0.050)(U-B) + (0.770 \pm 0.070)(B-V) \notag \\
&\quad + (0.720 \pm 0.040) \\
\label{equ:color transform2}
(g-r) &= (1.646 \pm 0.008)(V-R) - (0.139 \pm 0.004) \\
(r-i) &= (1.007 \pm 0.005)(R-I) - (0.236 \pm 0.003).
\label{equ:color transform3}
\end{align}

By applying the extinction correction, the colour evolution of SN\,2017ckj and the comparison with those of SNe IIb listed in Table~\ref{tab:SNe_IIb info} of Appendix~\ref{Sect:datainfo} is shown in Figure~\ref{Fig:color_evo}. 
Overall, the colour evolution of SN\,2017ckj is similar to that of typical SNe IIb.
The early colours of SN\,2017ckj exhibit a reddening trend, with the $B-V$ colour increasing by $\rm\sim1.5\,mag$, leading up to the peak at $\sim$50\,d.
This evolution is driven by the cooling of the expanding ejecta and the corresponding shift of the blackbody emission peak towards longer wavelengths.
After the red peaks, colours of $B-V$, $u-g$ and $g-r$ slowly decline over the next $\sim$150\,d.
This decline is expected as the ejecta becomes optically thin, allowing trapped photons to escape the inner ejecta.
However, there is no significant downward trend in the colours of $r-i$ and $i-z$ during 50 to 150\,d, which is due to the emergence of the [\Oi] $\lambda\lambda$6300, 6364 and [\Caii] $\rm\lambda\lambda$7292, 7324 lines that dominate the spectra at this phase.

It is worth noting that a rapid blueward colour evolution within about the first two weeks could be seen in SN\,2008ax and SN\,2020acat. 
The two SNe have been suggested to have a compact progenitor and lack a pronounced shock-cooling tail, which typically manifests as an initially very blue colour that subsequently transitions to red \citep{2022MNRAS.513.5540M}.
In contrast, the early colour evolution of SN\,2017ckj does not exhibit a rapid bluing phase in the first two weeks, consistent with the behaviour of most SNe IIb, including SN\,1993J.
These SNe generally possess a relatively extended H-rich envelope that expands rapidly after the explosion, resulting in a redder colour evolution until about 50 days.
This suggests that SN\,2017ckj may also have an extended envelope similar to those of typical SNe IIb.

\subsection{Pseudo-bolometric light curves}

A pseudo-bolometric light curve of SN\,2017ckj was constructed by using the photometric $uBgcVroiz$ bands.
The fitting procedure for the bolometric light curve employed the publicly accessible \texttt{Superbol}\footnote{\url{https://github.com/mnicholl/superbol}} programme \citep{2018RNAAS...2..230N}.
\texttt{Superbol} constructs the pseudo-bolometric light curves by integrating extinction-corrected fluxes across observed bands.
Taking into account the distance and reddening estimates reported in Section~\ref{sect:basic_inf}, we calculated the pseudo-bolometric light curve of SN\,2017ckj and several SNe IIb listed in Table~\ref{tab:SNe_IIb info} of Appendix~\ref{Sect:datainfo}, as shown in Figure~\ref{Fig:BL}.
To consistently compare luminosities, we constructed pseudo-bolometric light curves for our comparison sample of SNe IIb using the same available photometric bands.
If a SN lacked Sloan filters, we used the corresponding Johnson–Cousins filters to cover a similar wavelength range.
Note that SN\,2019tua lacked the $U$-band observation, and DES14X2fna had the photometric observation of $griz$ bands, which did not have a major effect on the comparisons.
Also shown in Figure~\ref{Fig:BL} is the theoretical decay slope of $\rm ^{56}Co$ (black dashed line), which is expected to dominate the late-time luminosity evolution of SNe.

The pseudo-bolometric light curve of SN\,2017ckj peaks at a luminosity of $L_{\text {peak}}=6.59_{-0.49}^{+0.49} \times 10^{42}\, \mathrm{erg}\,\mathrm{s}^{-1}$.
SN\,2017ckj displays a higher peak luminosity than the majority of SNe IIb shown in Figure~\ref{Fig:BL}, such as SN\,1993J, SN\,2015as, and SN\,2020acat.
Among the SNe of our sample, only SN\,2018gk and DES14X2fna have higher peak luminosity than SN\,2017ckj.
The late-time luminosity evolution of SN\,2017ckj resembles that of SN\,2011fu, although SN\,2017ckj does not exhibit a clear shock-cooling tail as seen in SN\,2011fu.
This similarity of late-time light curves suggests that a comparable mass of $^{56}$Ni was synthesised in the SN explosion.

\subsection{Bolometric light curve modelling} \label{blmodelling}

The late-time evolution of SN\,2017ckj exhibits a significantly steeper decline than that expected from the radioactive $\rm ^{56}Co$ decay (see Figure~\ref{Fig:BL}), which is likely due to an incomplete trapping of the $\gamma$-rays produced in the radioactive decay \citep{2015MNRAS.450.1295W}.
A simple relation to describe the late-time photometric evolution by \citet{1997ApJ...491..375C} for a sample of SE-SNe is:
\begin{equation}
L(t)=L_0(t) \times\left[1-e^{-\left(T_0 / t\right)^2}\right]
\end{equation}
Here, $T_0$ the full-trapping characteristic timescale defined as
\begin{equation}
T_0=\left(C \kappa_\gamma \frac{M_{\mathrm{ej}}^2}{E_k}\right)^{1 / 2}
\end{equation}
where $M_{\rm ej}$, $E_{k}$ and $\kappa_{\gamma}$ are the total ejected mass, kinetic energy, and the $\gamma$-ray opacity.
$C$ is a constant given by $C=(\eta-3)^2[8 \pi(\eta-1)(\eta-5)]$ for a density profile of the radioactive matter $\rho(r, t) \propto r^{-\eta}(t)$.

By assuming spherical symmetry and homologous expansion of shells with all radioactive material concentrated at the centre, \citet{2012A&A...546A..28J} provided the theoretical luminosity under the condition of complete trapping of the energy deposited by the decay of $\rm ^{56}Co$.
\begin{equation}
L_0(t)=9.92 \times 10^{41} \frac{M_{\mathrm{Ni}}}{0.07\, \mathrm{M}_{\odot}}\left(e^{-t / 111.4}-e^{-t / 8.8}\right) \mathrm{erg} \mathrm{~s}^{-1},
\end{equation}
where $M_{\mathrm{Ni}}$ is the $\rm ^{56}Ni$ mass expelled by the SN explosion. 

Therefore, we fitted the late-time bolometric light curve of SN\,2017ckj using MCMC simulations to obtain an estimate of the ejected $\rm ^{56}Ni$ mass and the full-trap characteristic timescale. 
The full bolometric light curve of SN\,2017ckj was also constructed using \texttt{Superbol} based on the blackbody assumption.
The best fitting $\rm ^{56}Ni$ mass is $\rm 0.21^{+0.05}_{-0.03} \,\rm M_\odot$ and the full-trapping characteristic timescale $T_0$ is $\rm 84.6^{+9.5}_{-12.1}\, d$.
The corresponding fitting and MCMC sampling results are presented in Figure~\ref{fig:latebolo fitting} of Appendix~\ref{sect:supfigure}.

\begin{figure}
   \centering
   \includegraphics[width = 1\linewidth]{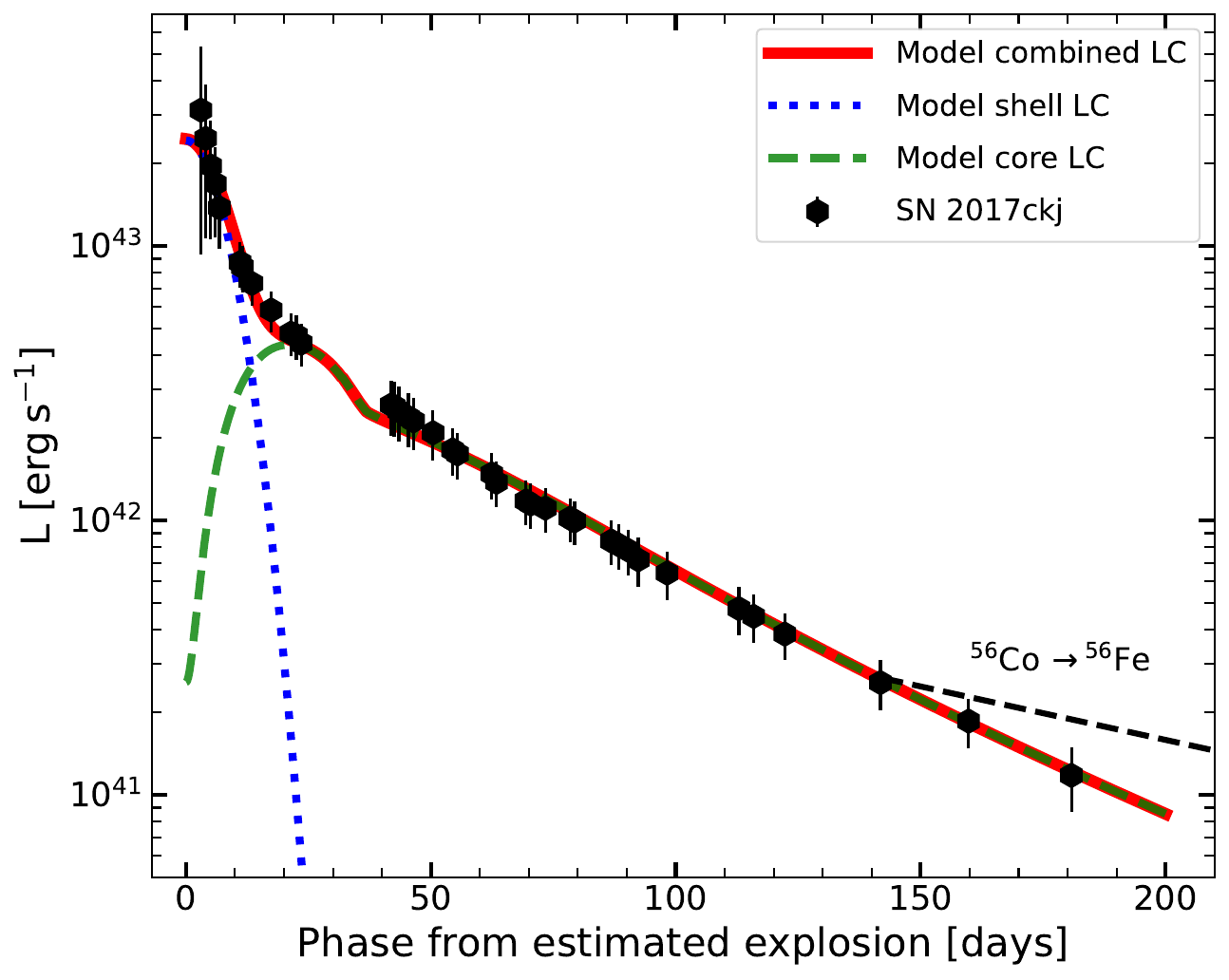}
   \caption{Full bolometric light curve of SN\,2017ckj (black dots) fitted with the two-component model of \citet{2016A&A...589A..53N}, including $\gamma$-ray leakage.
   The green and blue curves represent the contribution from the He-rich core and the extended H-envelope, respectively, while the red line shows the combined light curve.
   }
    \label{Fig:lcfitting}
\end{figure} 

Using the mean parameter from MCMC simulations, we applied the two-component light curve model of \citet{2016A&A...589A..53N} to fit the full bolometric light curve of SN\,2017ckj.
The model is based on a two-component configuration consisting of a dense inner He-rich region and an extended low-mass H-rich envelope for SNe IIb.
The light curve is thus the combination of radiation from the shock-heated ejecta and the radioactive decay of $^{56}{\rm Ni}$ to $^{56}{\rm Co}$.
Here, we adopted $\kappa = 0.3\, \mathrm{~cm}^2 / \mathrm{g}$ for the outer envelope and $\kappa = 0.2\, \mathrm{~cm}^2 / \mathrm{g}$ for the inner core, which is the same as the setting of \citet{2016A&A...589A..53N}.
In practice, we first fitted the core properties based on the previous late-time MCMC fitting results, and then constrained the envelope properties since the strength of the early-cooling emission depends on the underlying $^{56}{\rm Ni}$-heating light curve.

\begin{table}
    \centering
    \caption{Model parameters of SN\,2017ckj.}
    \renewcommand{\arraystretch}{1.25}  
    \begin{tabular}{l@{\hspace{1.mm}}c@{\hspace{1.mm}}c@{\hspace{1.5mm}}p{4.2cm}}
    \hline Parameters & Core & Envelope & Remarks \\
    \hline$R_0\ (10^{12} \mathrm{~cm})$ & 0.35 & 40 & Progenitor radius \\
    $M_{\mathrm{ej}}\ (M_{\odot})$ & 1.6 & 0.4 & Ejecta mass \\
    $M_{\mathrm{Ni}}\ (M_{\odot})$ & 0.21 & -  & Nickel mass \\
    $\kappa\ (\mathrm{~cm}^2 / \mathrm{g})$ & 0.2 & 0.3 & Thomson scattering opacity \\
    $E_{\text {tot }}\ (10^{51} \mathrm{erg})$ & 4.4 & 5.0 & Total energy \\
    $E_{\mathrm{kin}} / E_{\mathrm{th}}$ & 1.75 & 4.0  & The ratio of ejecta kinetic energy to thermal energy \\
    $A_g$ $\rm(day^2)$ & 7157 & 7157 & The factor represents the effectiveness of $\gamma$-ray trapping\\
    \hline
    \end{tabular}
    \label{tab:fitting parameters}
\end{table}

The fitting result of the bolometric light curve is shown in Figure~\ref{Fig:lcfitting}, and the best-fit parameters are summarised in Table~\ref{tab:fitting parameters}.
It is worth noting that $A_g = T_0^2$, where $A_g$ is the factor that represents the effectiveness of $\gamma$-ray trapping.
In Figure~\ref{Fig:lcfitting}, the high early-time blackbody temperature of SN\,2017ckj makes SED fitting with optical bands challenging, resulting in large uncertainties in the early luminosity.
Given the uncertainties of the early luminosity and the contribution from underlying $^{56}\mathrm{Ni}$ heating, we constrained the mass of the H-rich envelope to $M_{\rm ej} = 0.4^{+0.1}_{-0.1}\,\rm M_\odot$ to match the observed early high luminosity.

\begin{table}
    \centering
     \renewcommand{\arraystretch}{1.35}  
    \setlength{\tabcolsep}{1.5pt}  
    \caption{Model parameters of the comparison SNe IIb.}
    \begin{tabular}{ccccccccccc}
        \hline
         SNe IIb  & $M_{ \rm shell}$ & $M_{\rm Ni}$ & $M_{\rm core}$ & $E_{\rm kin }$  & References  \\
           & $(\rm M_\odot)$&$(\rm M_\odot)$ & $(\rm M_\odot)$& $(10^{51}\,\mathrm{erg})$ &  \\
        \hline
         2017ckj & $0.4^{+0.1}_{-0.1}$ & $0.21^{+0.05}_{-0.03}$ & 1.6 & 2.8 & This work \\
         1993J  & 0.1 & 0.1 & 2.15 & 2.4  & 1 \\
         2008ax  & - & 0.07-0.15 & 2-5 & 1-6 & 2 \\
         2011dh & 0.1 & 0.06 & $\sim2$ & 0.6-1.0 & 3 \\
         2011fu & 0.12 & 0.23 & 2.2 & 2.4 & 1 \\
         2015as  & 0.1 & 0.08 & 1.1-2.2 & 0.78 & 4 \\
         2020acat  & 0.1 & 0.12$\pm$0.03 & 2.3$\pm$0.4 & 1.2$\pm$0.3 & 5, 6 \\
         2021bxu  & $0.065^{+0.005}_{-0.005}$ &  $0.029^{+0.004}_{-0.005}$ & - & - & 7 \\
         2022crv  & 0.015-0.05 & 0.12$\pm$0.05 & 3.2-3.9 & 3.4 & 8 \\
        \hline
    \end{tabular}
    \begin{flushleft}
    References: 1. {\protect\cite{2016A&A...589A..53N}},
    2. {\protect\cite{2011MNRAS.413.2140T}},
    3. {\protect\cite{2012ApJ...757...31B}},
    4. {\protect\cite{2018MNRAS.476.3611G}},
    5. {\protect\cite{2022MNRAS.513.5540M}},
    6. {\protect\cite{2024A&A...683A.241E}},    
    7. {\protect\cite{2023MNRAS.524..767D}},
    8. {\protect\cite{2023ApJ...957..100G}}.
    \end{flushleft}
    \label{tab:SNe_IIb modelling parameter info}
\end{table}

We also summarised the model parameters of other SNe IIb in Table~\ref{tab:SNe_IIb modelling parameter info}.
SN\,1993J, SN\,2011dh, SN\,2011fu, and SN\,2015as possess relatively massive H-rich envelopes of $\sim0.1 \,\rm M_\odot$, whereas SN\,2021bxu and SN\,2022crv are inferred to have significantly lower mass envelopes, of the order of a few 0.01 M$_\odot$.
To account for the early high luminosity, SN\,2017ckj requires a more massive and extended H-rich envelope of $\sim 0.3-0.5\,\rm M_\odot$ to produce a luminous and long-lasting shock-cooling tail.
Although this value exceeds those of SNe IIb in our comparison sample, it still lies within the canonical mass range of SNe IIb \citep[$0.001-1.0\, \rm M_\odot$;][]{2017ApJ...840...10Y}.

Additionally, for SN\,2017ckj, the high $^{56}\mathrm{Ni}$ mass is particularly notable, exceeding that of the majority of SNe IIb.
Among known cases listed in Table~\ref{tab:SNe_IIb modelling parameter info}, only SN\,2011fu exhibits a comparable $^{56}\mathrm{Ni}$ mass of $0.23\,\rm M_\odot$ \citep{2016A&A...589A..53N}, and its late-time pseudo-bolometric light curve closely resembles that of SN\,2017ckj in Figure~\ref{Fig:BL}.
\citet{2016MNRAS.458.2973P} suggested the median $\rm ^{56}Ni$ mass for SNe IIb is approximately $0.11^{+0.04}_{-0.04}\,\rm M_\odot$.
\citet{2023ApJ...955...71R} also conducted a systematic study of the SE-SNe population and reported a mean $\rm ^{56}Ni$ mass of $\rm 0.066^{+0.006}_{-0.006}\,M\odot$ for SNe IIb.
However, the derived $\rm ^{56}Ni$ mass for SN\,2017ckj is significantly higher than both of these ranges.
The luminous SNe IIb DES14X2fna and SN\,2018gk would require $^{56}$Ni masses of $\sim 0.6$ and $\sim 0.4\,\rm M_\odot$, respectively, under the assumption of a standard $^{56}$Ni decay model \citep{2021MNRAS.505.3950G,2021MNRAS.503.3472B}.
Their exceptionally high luminosities may indicate the need for an additional power source, such as interaction with circumstellar material (CSM) or energy input from a rapidly rotating neutron star \citep[magnetar,][]{2021MNRAS.505.3950G}.

\begin{figure*}
   \centering
   \includegraphics[width = 0.9\textwidth]{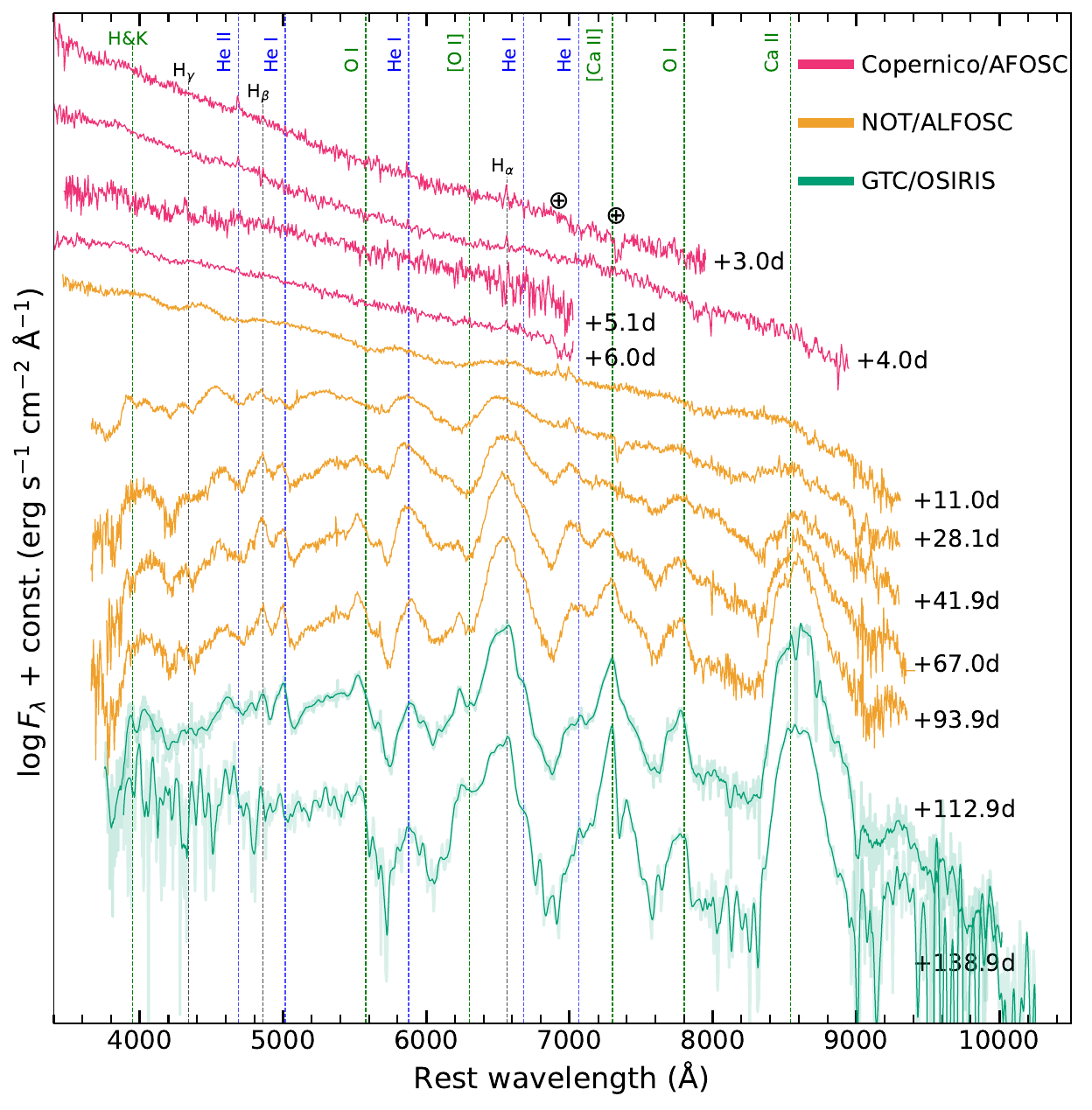}
   \caption{Spectral sequences of SN\,2017ckj.
    The positions of the principal transitions from H, He and other spectral elements are highlighted by the dashed vertical lines. 
    The $\oplus$ symbols mark the position of the strongest telluric absorption bands. 
    The phase of spectra based on the estimated explosion epoch (MJD~=~57837.1$\,\pm\,$0.1) is given on the right-hand side.
    All spectra have been corrected for redshift and extinction.
    The last two spectra, with lower signal-to-noise ratio (S/N), have been smoothed using a Savitzky-Golay filter.}
    \label{Fig:spec}
\end{figure*} 

\section{Spectroscopy}\label{sect:spectroscopy}

\subsection{Spectral sequence}

Figure~\ref{Fig:spec} shows the spectral evolution of SN\,2017ckj until the start of the nebular phase. 
The first spectrum of SN\,2017ckj is obtained on 28 March 2017 (MJD = 57840.1), approximately 3.0\,d after the estimated explosion date.
It exhibits a blue continuum with a blackbody temperature of $\sim \rm 22\,000\,K$, reflecting the high temperature of the ejecta shortly after the explosion. 
In addition, weak but distinct narrow emission lines of \ha, \hb, and \Heii $\lambda4686$ are also present at this phase. 
These flash-ionisation features are tentatively identified in the early-time spectra, and are thought to originate from the recombination of CSM that was ionised by the initial shock-breakout radiation pulse, similar to those observed in the early spectra of SN\,2018gk \citep{2021MNRAS.503.3472B}, SN\,2013cu \citep{2014Natur.509..471G}, and SN\,2022lxg \citep{2025A&A...700A.138C}.
By +6.0\,d, the blackbody temperature of the spectrum has cooled significantly, although it still shows a blue continuum, now corresponding to a blackbody temperature of around $\rm 10\,000\,K$.
At this stage, the \Heii $\lambda4686$ emission lines have faded, while the \ha line remains marginally visible.

At +11.0\,d from the explosion, the blue continuum has faded.
By +28.1\,d, broad P-Cygni profiles are more clearly developed, indicating an expanding photosphere and the transition into the photospheric phase.
Broad P-Cygni profiles of \ha, \hb, \hg, \Hei $\lambda\lambda5876,7065$, \Oi and \Caii begin to appear and steadily strengthen.
The \Hei $\lambda6678$ feature is not clearly detected, which may be due to its intrinsically low flux or significant blending with the prominent \ha emission, making it difficult to distinguish.
It is worth noting that the P-Cygni profile of \ha in the +41.9\,d spectrum exhibits a notch caused by the overlapping P-Cygni absorption of \Hei $\lambda$6678.
This notch feature is similar to that seen in SN\,1993J at +41\,d, although it is not as prominent as in SN\,1993J \citep{1995A&AS..110..513B}.

In the blue region of the optical spectrum at this stage, where \Feii\ features typically dominate, a strong absorption feature appears near $4900$\AA, comparable in depth to the \hb feature at +41.9\,d after the explosion.
This feature is generally attributed to \Feii $\lambda5018$.
Notably, the \Hei $\lambda$5016 could also contribute to the 4900\,\AA\ feature.
However, it is significantly weaker than \Hei $\lambda5876$, making it unlikely that \Hei $\lambda5016$ alone is responsible for the observed absorption \citep{2006sham.book..199D}.
The \Nii $\lambda$5005 could also contribute to this feature, but its contribution is likely limited due to the relative weakness of \Nii $\lambda$5680.

As time goes by, forbidden emission lines of [\Oi] $\lambda\lambda6300,6364$ and [\Caii] $\mathrm{\lambda\lambda}7291, 7324$, which are characteristic nebular-phase features, become prominent from +67.0\,d onwards.
In the nebular-phase spectra, the intensity of \ha profile does not show a significant weakening trend up to the last available spectrum on +138.9\,d, while the [\Oi] $\lambda\lambda6300,6364$ emission feature gradually strengthens over time.
In contrast, for SN\,1993J, the \ha emission profile is significantly weaker than [\Oi] emission in its +139 spectrum \citep{2000AJ....120.1487M}. 
This unusually strong \ha emission of SN\,2017ckj suggests the presence of a relatively massive hydrogen envelope of its progenitor.
The \Oi $\lambda7774$ feature appears at +28.1\,d and gradually strengthens in the subsequent spectra.

The \Caii NIR $\lambda\lambda8498, 8542, 8662$ triplet exhibits weak emission features as early as +11.0\,d, followed by rapid strengthening, ultimately becoming a significant feature in the nebular-phase spectra. 
In the red region of the late-time spectra, the \Oi $\lambda9263$ emission feature also becomes more prominent over time, despite the low S/N in that wavelength range.

\subsection{Flash ionisation in the early spectra}

\begin{figure}
   \centering
   \includegraphics[width = 1\linewidth]{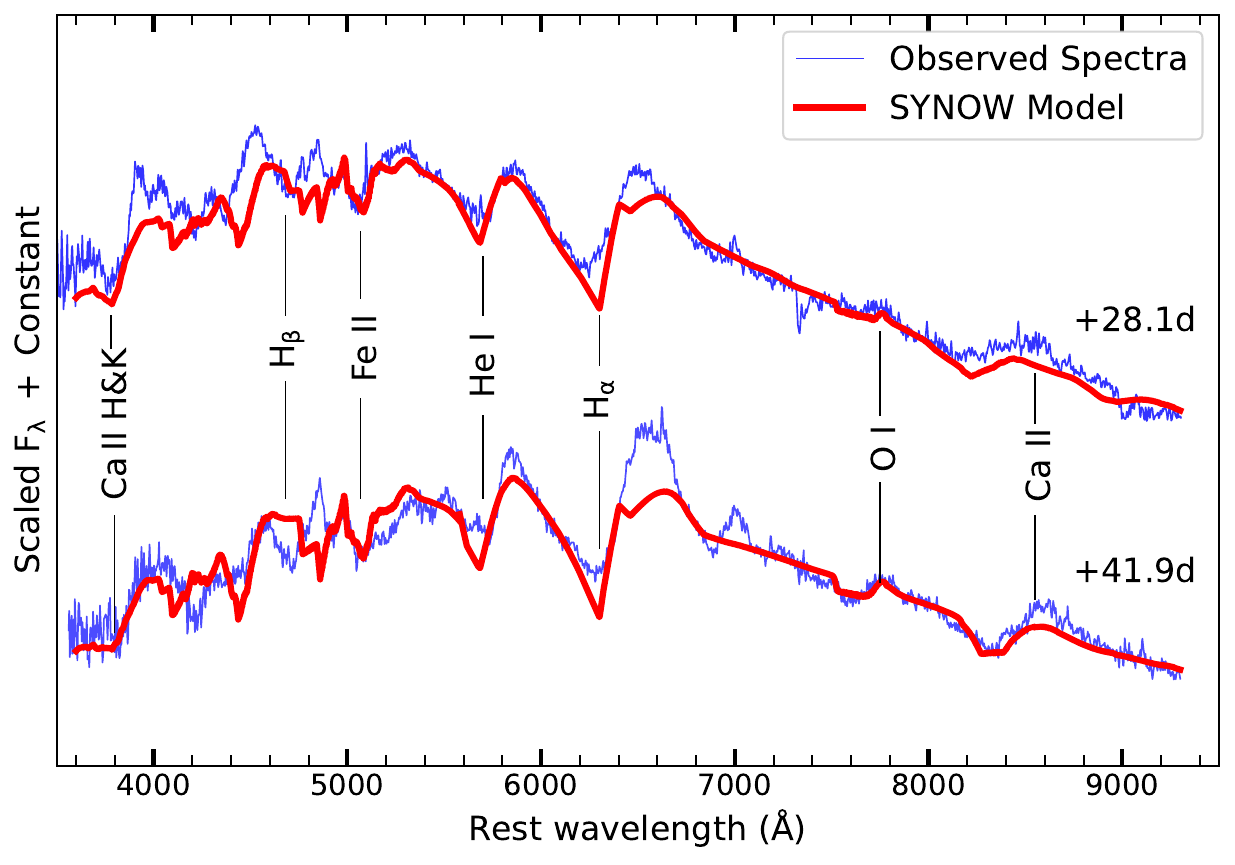}
   \caption{SYNOW models and line identifications for SN\,2017ckj at two epochs. 
   The observed spectra are corrected for extinction and redshift.
   All prominent permitted lines are labelled by marking the corresponding component.
   }
    \label{Fig:spec_model}
\end{figure}

In the first observed spectrum at +3.0\,d, the narrow emission line of \ha and \Heii likely originates from the recombination of CSM that was flash-ionised by the initial shock-breakout radiation pulse \citep[see e.g.][]{1985ApJ...289...52N,2014Natur.509..471G}.
The flash-ionised emission-line profiles of \ha and \Heii could be described by a combination of a broad Lorentzian and narrow Gaussian components \citep[e.g.][]{2017NatPh..13..510Y}.
However, for SN\,2017ckj, the flash-ionised emission components cannot be clearly decomposed due to the limited resolution of the early-time spectra.
The full width at half maximum (FWHM) of the overall \ha profile is approximately 700\,$\rm km\,s^{-1}$, while that of the \Heii profile is about 1100\,$\rm km\,s^{-1}$.

Assuming that the \ha emission arises from recombination in CSM photoionised by the shock breakout, we can estimate the progenitor mass-loss rate.
For this order-of-magnitude estimate, we neglect the additional complexities introduced by light travel time effects \citep{2019MNRAS.483.3762K}.
The \ha recombination luminosity of a fully ionised hydrogen wind is given by:
\begin{equation}
L_{\mathrm{H} \alpha}=\frac{\dot{M}^2 \alpha_{\mathrm{H} \alpha} \epsilon_{\mathrm{H} \alpha}}{4 \pi v_w^2 m_p^2 R_{\mathrm{in}}},
\end{equation}
where $\dot{M}$ is the mass-loss rate, $\alpha_{\mathrm{H} \alpha}$ is the Case B \ha recombination coefficient, representing the portion of the total Case B recombination coefficient that ultimately produces a \ha photon \citep{1995MNRAS.272...41S,2014AN....335..841T}, and $\epsilon_{\mathrm{H} \alpha} = 1.89\, \mathrm{eV}$ is the energy of a \ha photon.
Here, $v_w$ denotes the wind velocity, $m_p$ is the proton mass, and $R_{\mathrm{in}}$ represents the inner edge of the wind.

\citet{2021MNRAS.503.3472B} provided a simplified equation to estimate the required mass-loss rate:
\begin{equation}
\begin{aligned}
\dot{M} \approx & 1.4\left[\frac{L_{\mathrm{H} \alpha}}{10^{39} \mathrm{erg} \mathrm{~s}^{-1}} \cdot \frac{R_{\mathrm{in}}}{10^{14} \mathrm{~cm}}\right]^{1 / 2}\left[\frac{v_{\mathrm{w}}}{30 \mathrm{~km} \mathrm{~s}^{-1}}\right] \\
& \times 10^{-4}\ \mathrm{M}_{\odot}\, \mathrm{yr}^{-1}.
\end{aligned}
\end{equation}
For the first spectrum of SN\,2017ckj, the \ha luminosity is estimated to be $1.6 \times 10^{39}\,\mathrm{erg\,s^{-1}}$ based on a Gaussian profile fit to the \ha emission feature.
The inner edge of the wind is $R_{\mathrm{in}} \approx R_*+v_s t \approx 3.0 \times 10^{14} \mathrm{~cm}$ assuming a stellar radius of $R_* \sim 575\,R_\odot$ and a shock speed of $v_s = 10^4 \, \mathrm{km\,s^{-1}}$.
We adopted a typical wind velocity of $v_w = 30\, \mathrm{km\,s^{-1}}$ since the observed FWHM velocity is limited by the spectral resolution.
Under these assumptions, we estimated a mass-loss rate of $3.1\times10^{-4}\,\rm M_\odot\, \mathrm{yr}^{-1}$, which is similar to the typical mass-loss rates of SNe IIb \citep[$10^{-5}-10^{-4}\,\rm M_\odot\, \mathrm{yr}^{-1}$,][]{2014ARA&A..52..487S}.
The early spectra of SN\,2018gk also exhibit flash ionisation features, indicating a mass-loss rate of $2\times10^{-4}\,\rm M_\odot\, \mathrm{yr}^{-1}$ \citep{2021MNRAS.503.3472B}.
For comparison, \citet{2014ApJ...785...95M} suggested that the progenitor of SN\,2011dh had a mass-loss rate of $3\times10^{-6}\,\rm M_\odot\,\mathrm{yr}^{-1}$ during the final $\sim$1300 years before explosion.
The progenitor of SN\,2013df exhibited a mass-loss rate of $5.4\times10^{-5}\,\rm M_\odot\,\mathrm{yr}^{-1}$ \citep{2015ApJ...807...35M}.
Given the relatively high mass-loss rate of SN\,2017ckj, its progenitor may have undergone an unstable mass-loss phase before explosion, which could result in asymmetric distribution of the CSM density and inferred mass-loss rate around the SNe \citep[e.g.][]{1996ApJ...461..993F,2022ApJ...936...28T,2025Galax..13...72V}.

The second spectrum of SN\,2017ckj, obtained +4.0\,d after the explosion, reveals a narrow \ha emission feature with a luminosity of $L_{\ha} = 1.5\times 10^{39}\, \mathrm{erg\,s^{-1}}$, from which we infer a mass-loss rate of $3.4\times10^{-4}\,\rm M_\odot\, \mathrm{yr}^{-1}$.
The third spectrum could not be analysed due to insufficient S/N. 
The fourth spectrum, taken at +6.0\,d, similarly displays a narrow \ha feature with a luminosity of $L_{\ha} = 2.0\times 10^{39}\, \mathrm{erg\,s^{-1}}$, indicating a mass-loss rate of $4.7\times10^{-4}\,\rm M_\odot\, \mathrm{yr}^{-1}$.
Notably, the increase in the inferred mass-loss rate may be due either to a larger amount of CSM being excited by the shock or to the wind being further excited by photons from the ongoing circumstellar interaction.
Furthermore, a substantial fraction of the \ha narrow emission may originate from collisional excitation rather than recombination \citep{1996ApJ...461..993F}.
The model assumptions carry uncertainties, as we adopted a simplified prescription applied uniformly across different phases.


\subsection{Line identifications}

\begin{figure}
   \centering
   \includegraphics[width = 0.88\linewidth]{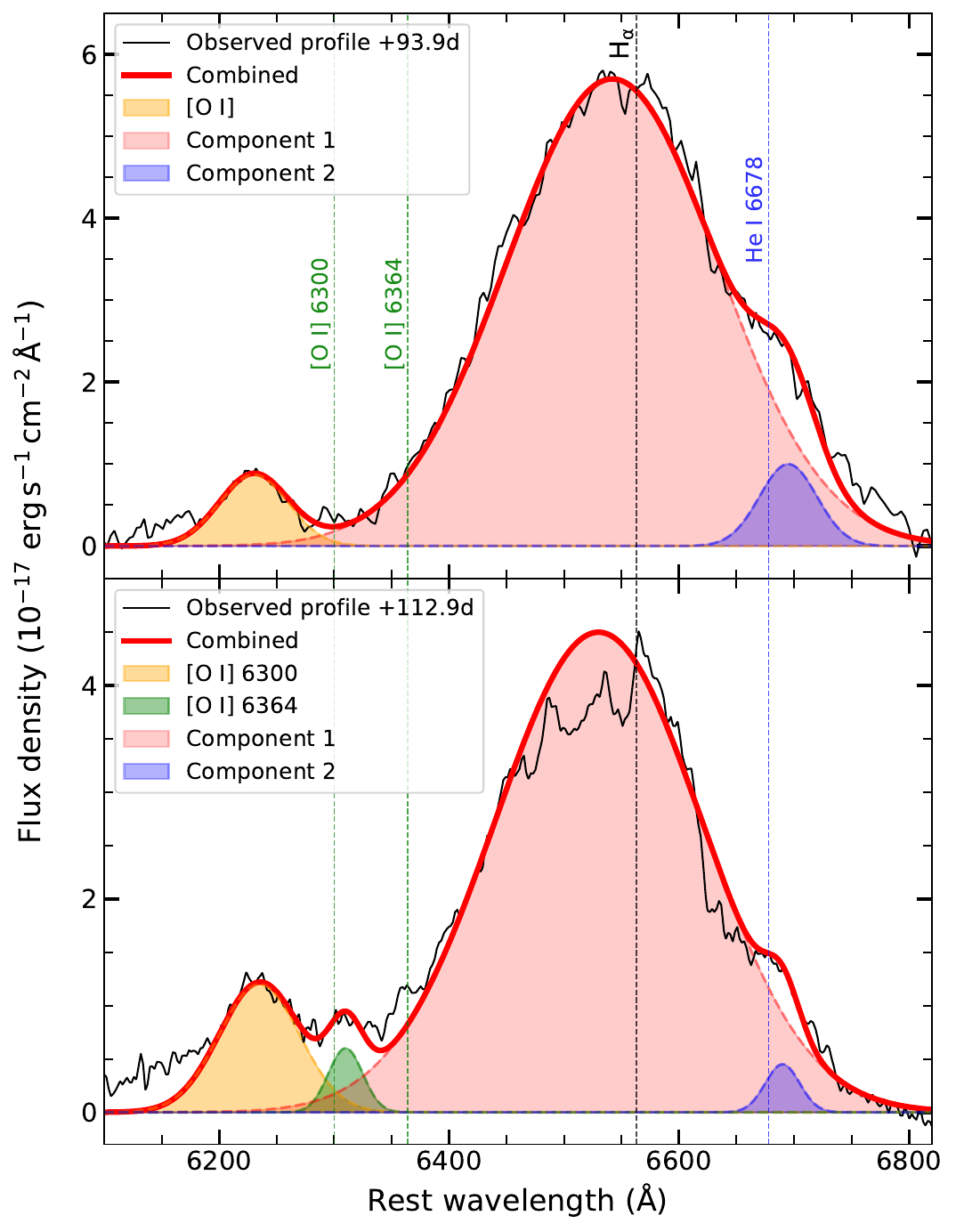}
   \caption{Decomposition of the emission profile around 6100-6800\,\AA\ in the +93.9\,d and +112.9\,d spectra. 
   The vertical dashed lines show the rest wavelengths of the [\Oi] doublet, \ha and \Hei.}
    \label{Fig:line_decomposition}
\end{figure} 

We modelled the spectra by adopting SYNOW \citep{1997ApJ...481L..89F,2002ApJ...566.1005B} and identified the spectral lines using a set of atomic species: \ha, \Hei, \Oi, \Feii, \Tiii, \Scii, \Caii, and \Baii.
The models with the line identifications at two phases are shown in Figure~\ref{Fig:spec_model}. 
The SYNOW code is based on several simplifying assumptions: spherical symmetry, homologous expansion, and the presence of a sharp photosphere that emits a blackbody continuum and is associated with a shock front at early times.
It is worth noting that SYNOW is suitable only for modelling spectra during the photospheric phase. 
SYNOW is designed to model permitted resonance lines formed via resonant scattering in rapidly expanding ejecta, and does not account for the low-density conditions that produce forbidden emission.
In our analysis, we modelled the +28.1 and 41.9\,d spectra and primarily focused on the absorption components of P-Cygni profiles, which offer insights into the expansion velocities of different line-forming layers, see Figure~\ref{Fig:spec_model}.
Although the SYNOW model does not fully reproduce these profiles, particularly on the blue side of the spectrum ($\lambda < 5500$\AA), the dominant line species are identified, such as \ha, \Hei, \Oi, \Caii, and \Feii.



Additionally, the late-time spectra of SN\,2017ckj exhibit the prominent emission profile around \ha line.
We therefore made the decomposition of the emission profile around 6100-6800$\text{\AA}$ in the late-time spectra, as shown in Figure~\ref{Fig:line_decomposition}.
The spectrum at +138.9\,d is not modelled due to its low S/N.
In the top panel, we used three Gaussian components to fit the full emission profile at +93.9\,d, including a blueshifted [\Oi], a broad and strong component 1, and a weak component 2.
The [\Oi] $\lambda6300$ profile is identifiable, whereas the $\lambda6364$ component appears to be weak or absent.
As a result, the [\Oi] $\lambda6364$ line was not included in the spectral fit at this phase.
Component 1 appears to be a distinct and broad \ha emission line, although it may represent a blended profile that includes contributions from \Cii $\lambda\lambda6578,6583$, \SiII $\lambda\lambda6347,6371$ and [\Nii] $\lambda\lambda6548,6583$.
Component 2, which is relatively weak, could be attributed to a redshifted \Hei emission feature.

In the bottom panel of Figure~\ref{Fig:line_decomposition}, the full emission profile at +112.9\,d is modelled using four Gaussian components.
Two of these represent the blueshifted [\Oi] $\lambda\lambda$6300, 6364 doublet, while the remaining two components account for additional blended flux.
At this stage, the [\Oi] doublet grows, with the $\lambda6364$ component appearing relatively distinct. 
The flux of Component 1 shows a gradual decrease.
Component 2 seems to exhibit a blueshifted absorption feature around 6650\,\AA.
Notably, both spectra display a noticeable bump on the blue side near the [\Oi] profile.
This feature, compared with the fitting results, indicates the presence of an underlying broad emission component.

\subsection{Comparison of type IIb SN spectra}

\begin{figure*}
   \centering
   \includegraphics[width = 0.9\textwidth]{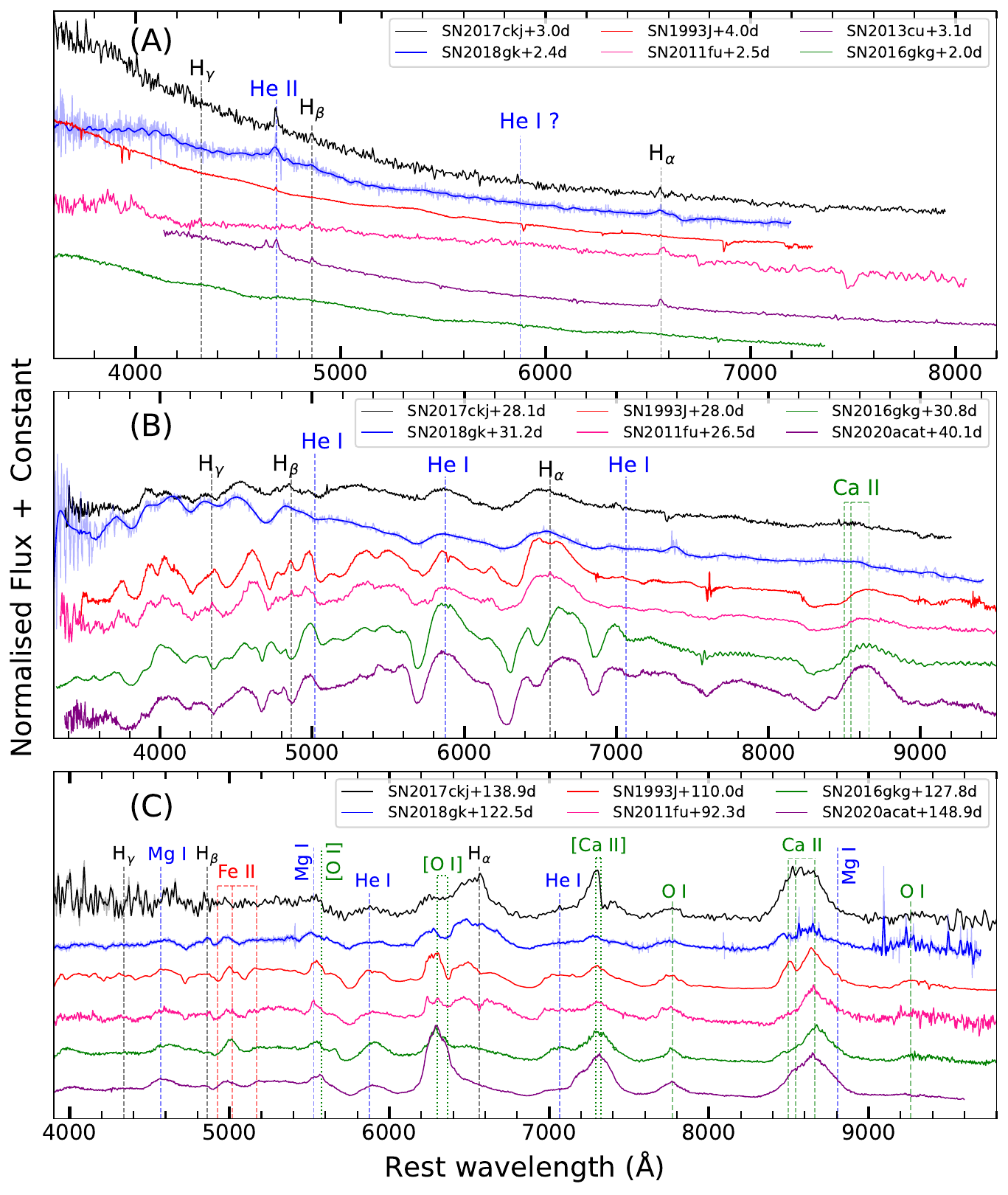}
   \caption{Early-, intermediate-, and late-time spectroscopic comparisons of SNe\,IIb.
   Phase from the explosion date is given in the legend and all spectra are shown in the rest frame.
   All spectra have been corrected for redshift and extinction.
   The key line features are marked, with allowed transitions indicated by dashed lines and forbidden lines represented by dotted lines.
   Some spectra with lower S/N have been smoothed using a Savitzky-Golay filter.
   }
    \label{Fig:spec_com}
\end{figure*} 

Figure~\ref{Fig:spec_com} compares the early-, intermediate-, and late-time spectra of SN\,2017ckj with those of other SNe IIb, including SN\,1993J, SN\,2011fu, SN\,2013cu, SN\,2016gkg, SN\,2018gk, and SN\,2020acat.
In panel~(A), we illustrate the earliest spectra of SN\,2017ckj on +3.0\,d with those of other SNe IIb at similar phases.
These early spectra exhibit a blue continuum, indicating a high blackbody temperature shortly after the explosion.
In addition, they display prominent emission features of \ha and \Heii $\rm \lambda 4686$. 
Notably, the early spectra of SN\,2017ckj likely exhibit a \Hei $\rm\lambda 5876$ emission feature, which is absent in other SNe IIb.

The spectrum of SN\,2017ckj at $\sim$30\,d post-explosion is also compared to the other SNe IIb at similar epochs, as shown in panel~(B) of Figure~\ref{Fig:spec_com}.
Although the \ha P-Cygni features are present in all spectra at this phase, their strength and width vary significantly among different SNe IIb.
For SN\,2017ckj, the \ha feature is quite broad, with a FWHM velocity of $\rm \sim12000\,km\,s^{-1}$, comparable to those of SN\,2018gk and SN\,2011fu.
The broad \Hei P-Cygni features in these SNe IIb exhibit noticeable blueshifts.
The central absorption feature within the \ha emission profile of SN\,2016gkg and SN\,2020acat is clearly visible and can be attributed to the P-Cygni absorption of the \Hei\,$\rm\lambda6678$ line, whereas this feature is absent in SN\,2017ckj and SN\,2018gk.
All spectra also show \Caii NIR P-Cygni features, although those in SN\,2017ckj and SN\,2018gk are relatively weak at this stage.
Overall, the spectrum of SN\,2017ckj at this stage closely resembles that of SN\,2018gk.

In panel~(C) of Figure~\ref{Fig:spec_com}, the early nebular spectrum of SN\,2017ckj at +138.9\,d is compared with a subset of SNe IIb.
At this phase, all the spectra are dominated by the prominent emission features of [\Oi] $\lambda\lambda6300,6364$, [\Caii] $\lambda\lambda7291,7324$, and \Caii NIR $\lambda\lambda8498, 8542, 8662$.
Additionally, other metal lines are present, including \Mgi] $\lambda4571$, \Feii $\lambda\lambda 4924,5018,5169$ and \Oi $\lambda\lambda7772,7774$.   
It is worth noting that the spectral features in the range of $6000-6800\, \text{\AA} $ for SN\,2017ckj at +138.9\,d differ from those of other SNe IIb. 
At this phase, SNe such as SN\,1993J and SN\,2011fu exhibit a strong [\Oi] $\lambda\lambda6300,6364$ emission profile, along with a distinct notch caused by blue-shifted \ha and \Hei lines.
However, in SN\,2017ckj, the \ha and \Hei lines are blended and cannot be separated into two distinct components.
In addition, the \Caii NIR $\lambda\lambda8498, 8542, 8662$ in SN\,2017ckj appears to be blended, similar to those in SN\,2011fu, SN\,2016gkg, and SN\,2020acat.
This contrasts with SN\,2018gk and SN\,1993J, where these emission features can be clearly resolved into multiple components.
The observed blending profile may originate from an aspherical explosion geometry, as suggested by \citet{2005Sci...308.1284M}.

\subsection{Spectra line profile and velocity evolution} 

\begin{figure*}[t]
    \centering
    \begin{minipage}{0.195\textwidth}
        \centering
        \includegraphics[width=\textwidth]{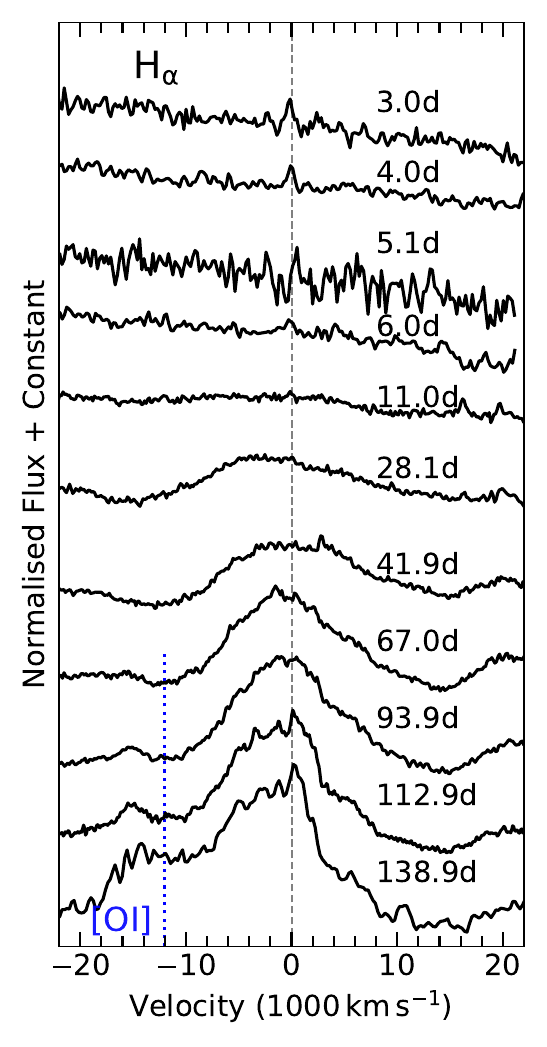}
    \end{minipage}%
    \begin{minipage}{0.195\textwidth}
        \centering
        \includegraphics[width=\textwidth]{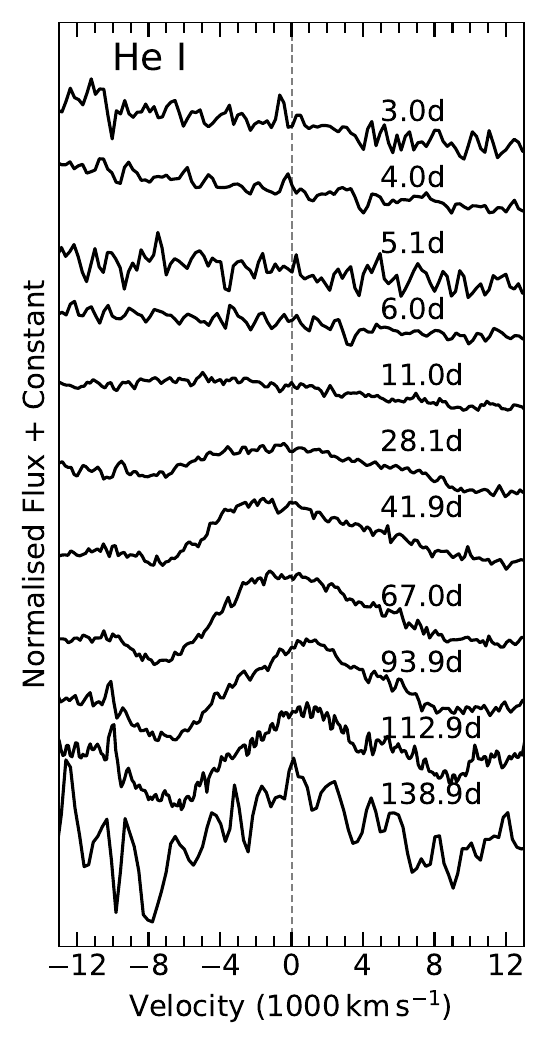}
    \end{minipage}%
    \begin{minipage}{0.195\textwidth}
        \centering
        \includegraphics[width=\textwidth]{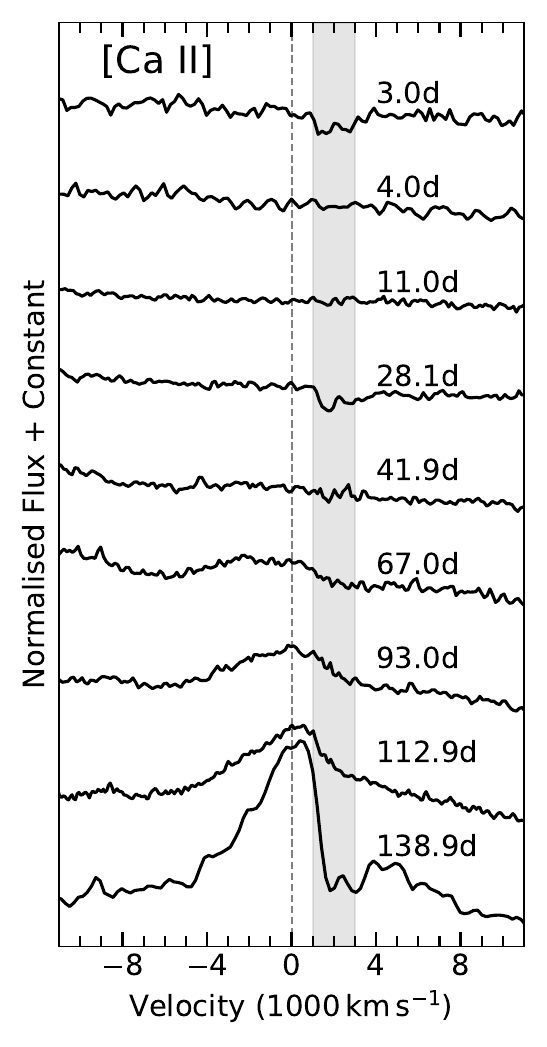}
    \end{minipage}%
    \begin{minipage}{0.195\textwidth}
        \centering
        \includegraphics[width=\textwidth]{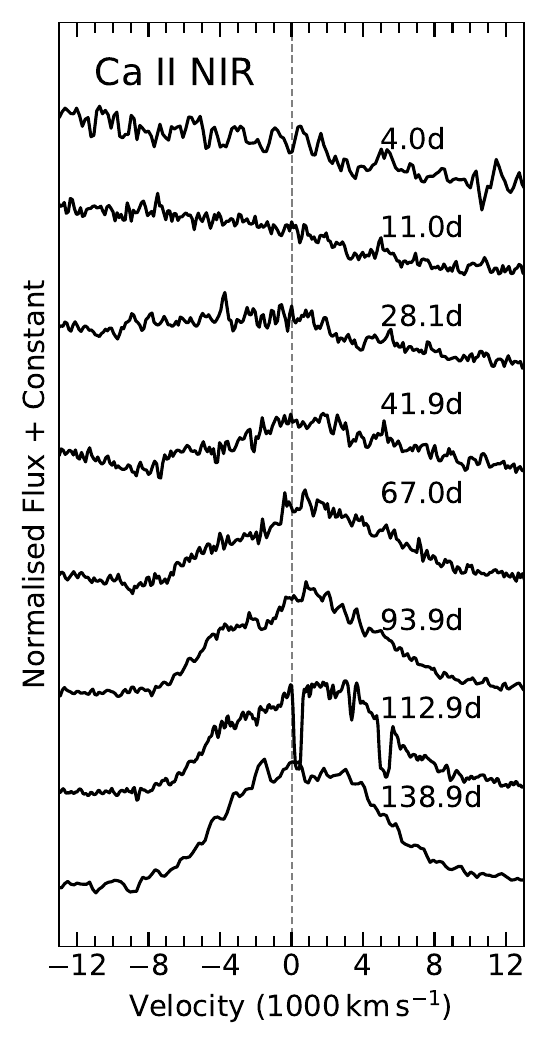}
        \end{minipage}
    \begin{minipage}{0.195\textwidth}
        \centering
        \includegraphics[width=\textwidth]{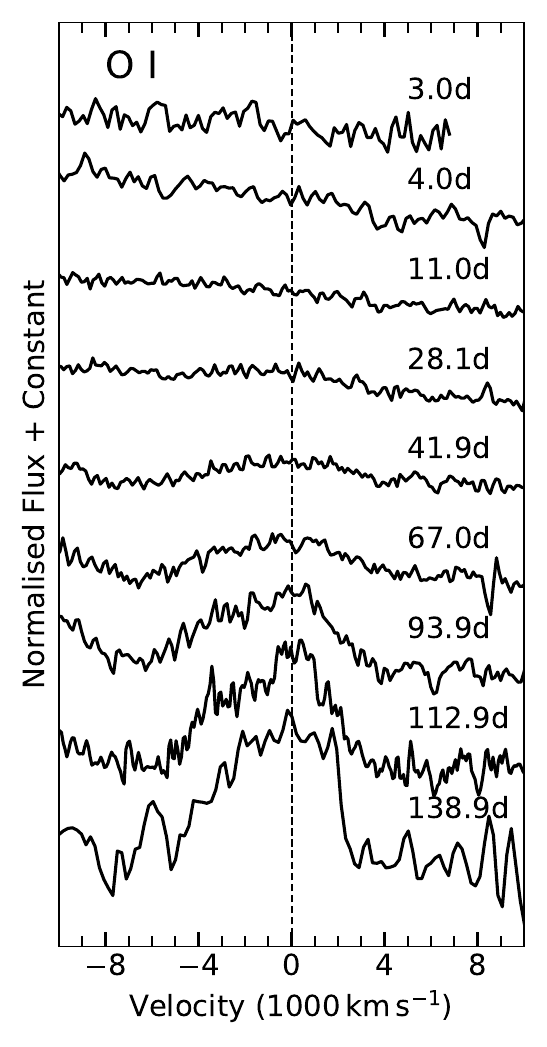}
    \end{minipage}
    \caption{Line profiles of \ha, \Hei, $\rm [\Caii]$, $\rm \Caii\ NIR$, and \Oi within the spectra of SN\,2017ckj.
    The dashed lines represent the emission velocities corresponding to 6563, 5876, 7291, 8571, and 7774\,\AA\ lines. 
    The epoch of each spectrum is given in the rest frame, relative to the estimated explosion date.}
    \label{fig:line_profiles}
\end{figure*}

\begin{figure}
   \centering
   \includegraphics[width = 0.9\linewidth]{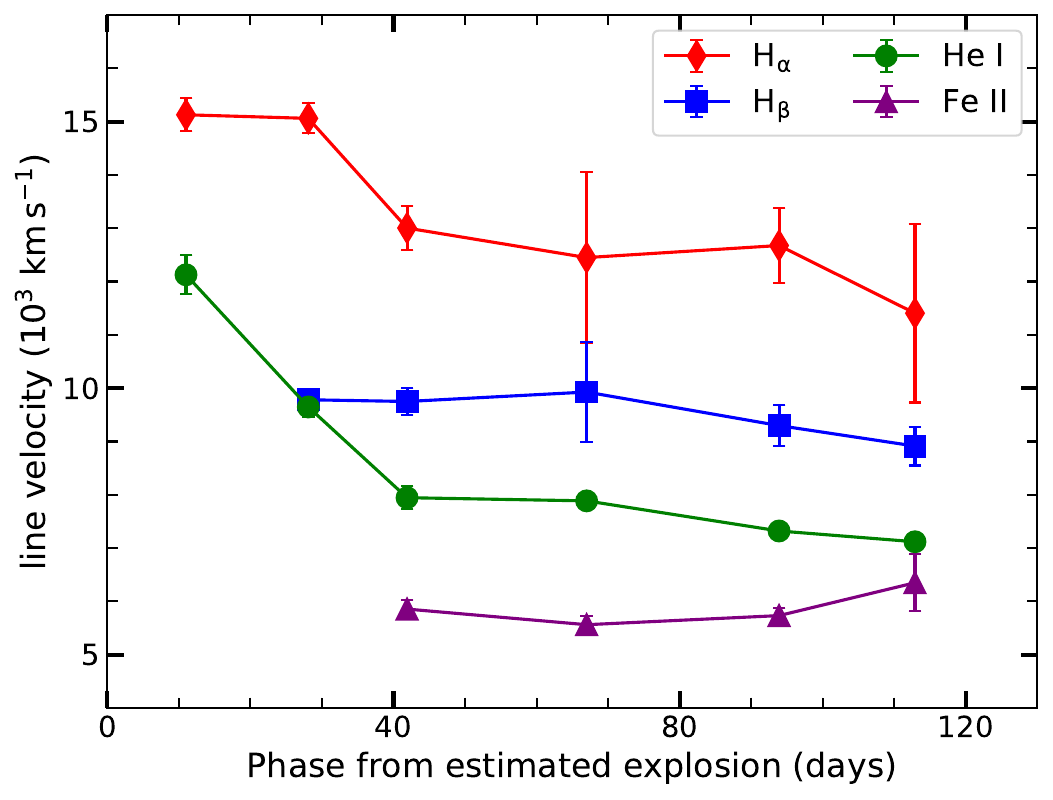}
   \caption{Line velocity evolution of \ha, $\rm H_\beta$, \Hei~$\lambda\,5876$, and \Feii~$\lambda\,5018$ for SN\,2017ckj, derived from the troughs of their P-Cygni profiles.}
    \label{Fig:line_velocity}
\end{figure}

The evolution of the profiles of the \ha, \Hei, \Caii NIR, and \Oi peaks in the spectra of SN\,2017ckj are shown in Figure~\ref{fig:line_profiles}.
The early spectrum at +3.0\,d exhibits narrow flash-ionised \ha and \Hei emission lines.
The spectrum at +28.1\,d displays distinct P-Cygni profiles, including \ha, \Hei, \Caii and \Oi, with significantly higher velocities, indicating a rapidly expanding envelope.

In the panel showing the \ha profile (Figure~\ref{fig:line_profiles}, left), the blue dotted line indicates the position corresponding to [\Oi] $\lambda6300$ in the \ha velocity coordinate.
The late-time spectrum at +93.9\,d shows a weak [\Oi] $\lambda\lambda6300,6364$ emission profile which gradually strengthens.
In the nebular spectra of SNe IIb, the [\Oi] emission profile gradually strengthens over time and becomes prominent at late phases, as seen in the late-time spectrum of SN\,2020acat and SN\,2016gkg in Figure~\ref{Fig:spec_com}.
This is driven by radioactive decay energy deposited in the oxygen-rich core, now revealed by the receding photosphere, allowing forbidden line emissions to emerge in the low-density and expanding ejecta.

At late times, the [\Caii] $\lambda\lambda$7291, 7324 lines exhibit a notable asymmetry, while the \Caii NIR triplet remains largely symmetric.
It is worth noting that distinct absorption features (grey region) appear on the red side of the [\Caii] emission profile, particularly in the spectra obtained at +3.0, +28.1, and +138.9 days. 
These absorption features are likely the result of dust extinction, which contributes to the asymmetric structure observed in the [\Caii] emission in the latest spectrum.
Recently, \citet{2025A&A...698A.293D} proposed that strong forbidden lines in noninteracting SNe II, such as [\Oi] and [\Caii], are significantly influenced by dust located within velocities of \textless$\rm 2000\,km\,s^{-1}$.
Additionally, the \Oi $\lambda$7774 profile of the last spectrum appears to exhibit a boxy feature, deviating from the Gaussian-like profile observed at earlier phases, although the spectrum has low S/N.
Meanwhile, the peak of \ha profile shows a similar but weaker flattening feature.
This \Oi and \ha flat-topped profile may result from dust absorption on the red side of the emissions or CSM interaction.

The velocity evolution of spectral lines, including \ha, \hb, \Hei\ $\lambda5876$, and \Feii\ $\lambda5018$, is shown in Figure~\ref{Fig:line_velocity}.
The line velocities are determined from the absorption minima of P-Cygni profiles, using MCMC-based fitting with a two-component Gaussian profile.
The early spectra exhibit a blue continuum, while broad P-Cygni profiles begin to emerge from +11.0\,d.
The early \ha and \Hei exhibit high expansion velocities of approximately $\rm15000\,km\,s^{-1}$ and $\rm12000\,km\,s^{-1}$, respectively.
Over the first epoch to +41.9\,d, the \ha and \Hei velocities decline rapidly. 
From +41.9\,d to +112.9\,d, the \ha, \hb, \Hei, and \Feii velocities show an approximately constant plateau.
At this stage, the velocities of \ha\ and \hb\ remain nearly constant at $\rm\sim 13000\,km\,s^{-1}$ and $\sim\rm10000\,km\,s^{-1}$, respectively.
In contrast, \Hei\ and \Feii\ maintain lower velocities of $\sim\rm 7500\,km\,s^{-1}$ and $\sim\rm 6000\,km\,s^{-1}$.

Overall, the line velocities of absorption features identified in SN\,2017ckj are consistent with those of typical SNe IIb \citep[see][]{2022MNRAS.513.5540M}.
The absorption velocity of \ha ($\rm 12500\pm800\,km\,s^{-1}$), derived from the weighted measurements of a sample of SNe IIb by \citet{2016ApJ...827...90L}, matches well with SN\,2017ckj.
Similarly, its \Hei velocity agrees with the weighted mean of SE-SNe samples \citep[$\rm 8000\pm500\,km\,s^{-1}$,][]{2018A&A...618A..37F}.
In contrast, the \Feii velocity in SN\,2017ckj is notably lower than most SNe IIb \citep[$\rm 8400\pm1400\,km\,s^{-1}$,][]{2016ApJ...827...90L}, but comparable to SN\,1993J ($\sim\rm6000\,km\,s^{-1}$) and SN\,2021bxu ($\sim\rm5000\,km\,s^{-1}$).
On the other hand, the velocities of \ha and \hb are higher than those of \Hei, which in turn are higher than those of \Feii.
This trend is consistent with the expected layered structure of the progenitor star, where heavier elements are concentrated closer to the core with lower velocities.

\section{Discussion} \label{sect:discussion}

\subsection{The nature of light curve evolution}

A prominent shock-cooling tail is observed in many SNe IIb, including SN\,1993J \citep{1994AJ....107.1022R}, SN\,2011fu \citep{2015MNRAS.454...95M}, SN\,2016gkg \citep{2018Natur.554..497B}, SN\,2024uwq \citep{2025arXiv250502908S}, and SN\,2024aecx \citep{2025arXiv250519831Z}.
This early-time shock-cooling phase arises from the rapid cooling of the outer stellar layers after the shock breakout at the surface.
However, there are some SNe IIb that do not show two distinct peaks and shock-cooling tail, such as SN\,2008ax \citep{2008MNRAS.389..955P}, SN\,2020acat \citep{2022MNRAS.513.5540M}, SN\,2022crv \citep{2023ApJ...957..100G}, and SN\,2024abfo \citep{2025A&A...698A.129R}.
The absence of an initial long-duration shock breakout cooling peak and distinct shock-cooling tail implies that the progenitor was relatively compact, typically lacking an extended hydrogen envelope.
In the light curves of SN\,2017ckj, there is a short interval between the initial multi-band observation and the last non-detection limit in $c$ band.
Due to the scarcity of data in this early phase, the possibility of a fast-evolving shock-cooling peak occurring during this interval cannot be ruled out.

The majority of SNe IIb that exhibit a distinct shock-cooling tail are believed to originate from progenitors with an extended hydrogen envelope of approximately 0.1$\rm M_\odot$ \citep[e.g.][]{2016A&A...589A..53N}.
In contrast, those lacking a clear shock-cooling tail are thought to arise from more compact progenitors with thinner hydrogen envelopes, typically less than 0.1$\rm M_\odot$.
For example, SN\,2008ax does not exhibit a shock-cooling tail, and its \ha feature disappears by $\sim$50 days after the explosion.
Therefore, it was suggested that its progenitor possesses a hydrogen envelope of approximately several times 0.01$\rm M_\odot$  \citep{2011ApJ...739...41C}.
Similarly, SN\,2022crv was considered to have a 0.05$\,\rm M_\odot$ hydrogen envelope, placing it as a continuum between type Ib and type IIb SNe \citep{2023ApJ...957..100G,2024ApJ...974..316D}.
The \ha feature of SN\,2022crv disappears around 35 days after the explosion.

A subset of SNe IIb with peculiar light curve evolution has been identified, including SN\,2018gk, DES14X2fna, and SN\,2017ckj.
As shown in Figure~\ref{fig:Abs_V}, these three SNe IIb exhibit higher peak magnitudes compared to the majority of SNe IIb ($\sim 1-2\,$mag in the $V$ band).
Their light curve evolution deviates from that of typical SNe IIb, including both those with and without an early shock-cooling tail.
Under the assumption of a standard $^{56}$Ni decay model, DES14X2fna and SN\,2018gk require unusually large $^{56}$Ni masses of $\sim$0.6 and $\sim$0.4 $\rm M_\odot$, respectively \citep{2021MNRAS.505.3950G,2021MNRAS.503.3472B}.
Such high luminosities may suggest the presence of an additional power source, such as interaction with CSM or energy input from a rapidly rotating magnetar \citep{2021MNRAS.505.3950G}.
Additionally, SN\,2022lxg appears to be a SE transitional object, exhibiting early-time spectra similar to those of SNe IIb, while later resembling an interacting type II supernova \citep{2025A&A...700A.138C}.
Its light curve evolution is comparable to that of SN\,2017ckj, characterised by a luminous peak ($\rm M_g=-19.41\,$mag) and a short rise time of $\lesssim$10 days.
Notably, the early-time $g$- and $r$-band light curves of SN\,2022lxg exhibit bumps likely caused by CSM interaction, and the late spectral features were suppressed because of CSM interaction after $\sim$35 days.
\citet{2025A&A...700A.138C} suggested that the interaction between the ejecta and the CSM might contribute to the early-time luminous peak of SN\,2022lxg and potentially blend the two components of the SNe IIb light curve.  

For SN\,2017ckj, a $^{56}$Ni mass of $\sim\rm 0.21\,\rm M_\odot$ is required to explain its late-time bolometric luminosity evolution. 
This value is not particularly large compared to those of DES14X2fna and SN\,2018gk.
The early spectra exhibit the flash ionisation features, similar to SN\,2018gk. 
In comparison with typical SNe IIb, SN\,2017ckj exhibits a long-lasting and prominent \ha emission profile, which persists up to the last available spectrum at $\rm +138.9\,d$.
By contrast, other SNe IIb generally do not show such a distinct \ha emission feature at similarly late phases.
Combining these observational characteristics, we therefore suggested that the peculiar light curve of SN\,2017ckj may result from the following scenarios: 
(1) A progenitor with a relatively massive and extended H-rich envelope (i.e. the outer layers were not significantly stripped) gives rise to a long-lasting shock-cooling phase after the explosion that overlaps and blends with the subsequent $^{56}$Ni-powered peak.
This result of overlap leads to a relatively rapid rise in the light curve.
This scenario is consistent with the bolometric light curve modelling results shown in Figure~\ref{Fig:lcfitting}; 
(2) A rapidly evolving shock-cooling peak has occurred before the earliest multi-band observations, though it is not seen in the previous ATLAS survey.
The subsequent luminous and fast peak could result from a combination of CSM interaction and $^{56}$Ni decay.
The shock-cooling tail of SNe IIb typically lasts for approximately 1–10 days, depending on the properties of the envelope and the degree of $\rm^{56}Ni$ mixing.
For SN\,2016gkg, the shock-cooling tail persists for about 4 days in the $V$ band \citep{2017ApJ...836L..12T}, whereas for SN\,2011fu it lasts roughly 10 days \citep{2015MNRAS.454...95M}.
Due to the photometric limits of the ATLAS survey and the $\sim$4 days interval between the last non-detection and the first detection, we cannot rule out the presence of a faint, rapidly evolving shock-cooling phase prior to the earliest multi-band observations.
This scenario is consistent with the presence of CSM, as indicated by the flash ionisation features observed in the early-time spectra.
Additionally, the CSM interaction may suppress the emergence of other spectral features, as observed in SN\,2022lxg, whose spectrum around 80\,d is dominated by \ha and \Caii lines, with most other features being notably weak or absent.
In contrast, the late-time spectrum of SN\,2017ckj at +138.9\,d exhibits more prominent spectral features, which may suggest the presence of a relatively thin CSM shell that was rapidly overtaken by the ejecta and thus influenced only the early-time spectra.
Further late-time, high-quality spectroscopy and multi-wavelength observations are required to better constrain the nature and extent of the CSM interaction.

\subsection{The progenitor parameters of SN 2017ckj}

\begin{figure}
   \centering
   \includegraphics[width = 1\linewidth]{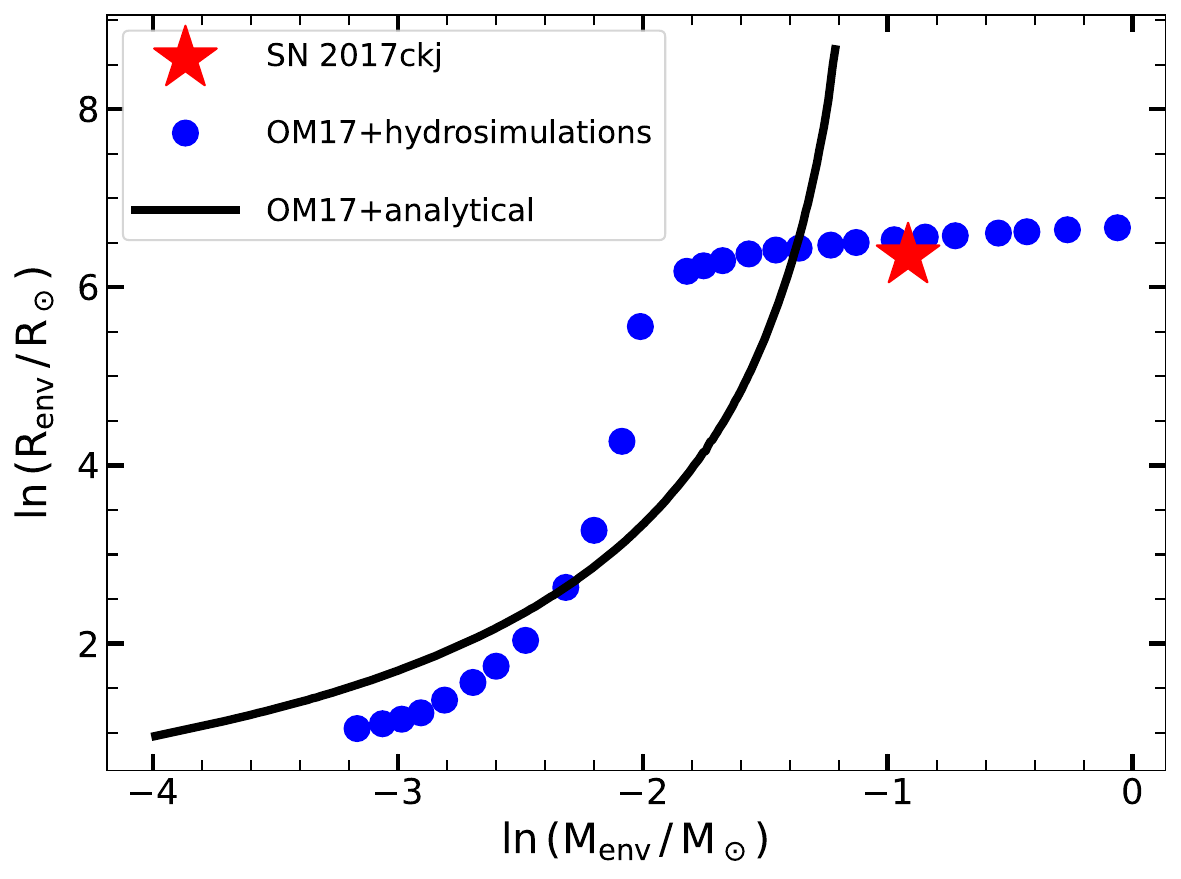}
   \caption{
   Relation between the radius and mass of the envelope for the SNe IIb progenitor models (blue dots) from \citet{2017ApJ...840...90O}.
   The black line represents the analytical relation derived by \citet{2017ApJ...840...90O}, which captures the properties of the numerical evolution models.
   The parameters derived for SN\,2017ckj are indicated by the red star.
   This figure is modified from \citet{2023ApJ...957..100G}.
   }
    \label{Fig:M-R}
\end{figure}

In Section~\ref{blmodelling}, we proposed a progenitor model for SN\,2017ckj consisting of a massive H-rich envelope of $\sim\rm 0.4\,\rm M_\odot$ and an extended envelope radius of $\rm \sim 575\,R_\odot$ by adopting the two-component model of \citet{2016A&A...589A..53N}.
\citet[][]{2017ApJ...840...90O} calculated a grid of binary evolution models for SNe IIb and provided a sequence to explore the progenitor properties of SNe IIb.
Figure~\ref{Fig:M-R} shows the relations between stellar radius and H-rich envelope mass, derived numerically from evolutionary models (blue points) and analytically assuming a radiative envelope (black line) under dynamical and thermal equilibrium.
To compare with their results, we over-plotted the derived envelope parameters for SN\,2017ckj (red star).
The envelope parameters of SN\,2017ckj are consistent with the relation inferred from hydrodynamical simulations, despite the uncertainty in the envelope mass.
Note that the discrepancy between the two approaches becomes more pronounced as the envelope mass increases and convection becomes more significant \citep{2017ApJ...840...90O}.

We constructed the bolometric luminosity and derived a high $\rm ^{56}Ni$ mass of $\rm 0.21^{+0.05}_{-0.03} \,\rm M_\odot$. 
Notably, SN\,2011fu exhibits a comparable late-time bolometric luminosity evolution to SN\,2017ckj, with an estimated $\rm ^{56}Ni$ mass of $\rm 0.23\,\rm M_\odot$ \citep{2016A&A...589A..53N}.
\citet{2013MNRAS.431..308K} also fitted the light curve of SN\,2011fu with the analytic model of \citet{1989ApJ...340..396A} and suggested a $\rm ^{56}Ni$ mass of $\rm 0.21\,\rm M_\odot$.
Therefore, a $\rm ^{56}Ni$ mass of $\rm 0.22\,\rm M_\odot$ is plausible for SN\,2017ckj, although both SN\,2017ckj and SN\,2011fu show higher $\rm ^{56}Ni$ masses compared to SNe IIb population \citep[$\rm .066^{+0.006}_{-0.006}\,\rm M_\odot$, ][]{2023ApJ...955...71R}.
Recent studies have reported two luminous SNe IIb, SN\,2018gk and DES14X2fna, with high $\rm ^{56}Ni$ mass of $\sim 0.6$ and $0.4\,\rm M_\odot$, respectively, based on the assumption of the standard $\rm ^{56}Ni$ decay model \citep{2021MNRAS.505.3950G,2021MNRAS.503.3472B}.
However, some studies suggested that the $^{56}$Ni yield in typical CCSNe cannot be too high \citep[$\sim0.2\,\rm M_\odot$;][]{2011ApJ...741...97D,2020ApJ...890...51E}.
\citet{2024A&A...682A.123T} considered that convective-core overshooting plays an important role in element production (especially $\rm ^{56}Ni$) during the final evolutionary stages of massive stars.
With an appropriate overshooting prescription, they were able to account for $^{56}$Ni mass yields of up to $0.26\,\rm M_\odot$.
However, for overluminous SNe IIb events, such as SN\,2018gk and DES14X2fna, additional energy sources beyond radioactive decay are likely required.

\citet{1989ApJ...343..323F} proposed that the flux ratio of the [\Caii] $\lambda\lambda$7291, 7324 to the [\Oi] $\lambda\lambda$6300, 6364 emission lines provides a useful diagnostic for estimating the progenitor mass on the main-sequence stage.
The [\Oi] $\lambda\lambda$6300, 6364 emission could trace the oxygen mass synthesised in the core, which reflects the progenitor’s initial mass, whereas the calcium emission primarily originates from explosive nucleosynthesis and is thus largely independent of the progenitor’s initial mass \citep[e.g.][]{1995ApJS..101..181W,1996ApJ...460..408T}.
\citet{1987ApJ...322L..15F,1989ApJ...343..323F} suggested that the forbidden emission line ratio [\Caii]/[\Oi] is expected to show an almost constant value at late epochs.
\citet{2004A&A...426..963E} examined the late-time evolution of the ratio and showed that the ratio with that of a sample of SNe Ib/Ic is stable after $\sim$250 days of the explosion, consistent with theoretical expectations.
Additionally, \citet{2014MNRAS.439.3694J} provided the late-time observed luminosity evolution of [\Oi], \Mgi] and \Nai D with different initial progenitor masses, which can be used to constrain the oxygen mass.
For SN\,2017ckj, the last spectrum was obtained at +138.9\,d after the explosion.
A distinct and strong [\Oi] emission feature was not observed in the last spectrum, and the prominent \ha-like emission makes it difficult to measure flux accurately.
In addition, the [\Caii] emission profile appears to be affected by dust absorption.
To obtain a rough estimate, the forbidden emission line ratio [\Caii]/[\Oi] is $\sim$1.4, suggesting that the progenitor of SN\,2017ckj has an initial mass of $\rm \sim15-17\,\rm M_\odot$ \citep{2014MNRAS.439.3694J,2024A&A...687L..20F}.
Notably, the measurements of both [\Caii] and [\Oi] have large uncertainties, and this result should be treated with caution.

\section{Concluding remarks} \label{sect:remarks}

We presented the discovery and follow-up observations of the luminous SN\,2017ckj, a fast-rising SN IIb with a $V$-band rise time of $\rm \sim 5.0$\,d, which is shorter than that of typical SNe IIb.
Its absolute fitted peak magnitude ($M_{\mathrm{V}}=-18.49\pm0.18\,\mathrm{mag}$) is also brighter than the majority of SNe IIb ($\rm\sim -17.5\,mag$), apart from overluminous SN\,2018gk and DES14X2fna.
After the explosion, the light curves in the $Vriz$ bands show a linear decline after peak, and the other bands follow a similar trend starting approximately 40 days after their respective peaks.
Spectroscopic monitoring of SN\,2017ckj spanned $\sim$140 days, providing a comprehensive optical dataset from early light to the onset of the nebular phase.
The earliest spectrum, obtained at +3.0\,d, exhibits flash-ionised features of \ha and \Heii, along with the blue featureless continuum.
During the intermediate phase, the spectra develop prominent P-Cygni profiles of \ha and \Hei, similar to SN\,2018gk.
The late-time spectrum at +138.9\,d displays a strong \ha-like emission profile while [\Oi] feature remains weak.
In comparison, in the spectra of prototypical SN\,1993J, the \ha feature weakens and becomes fainter than [\Oi] features at the same stage.
Unfortunately, SN\,2017ckj became too faint to be observed further due to its large distance of $158\pm11\, \mathrm{Mpc}$.
To explain the peculiar light curve evolution and the late-time distinct \ha profile at the last +138.9\,d spectrum, we considered that the progenitor of SN\,2017ckj likely possesses a massive H-rich envelope of $0.4^{+0.1}_{-0.1}\,\rm M_\odot$ and a high $\rm ^{56}Ni$ mass of $0.21^{+0.05}_{-0.03} \,\rm M_\odot$.

Several ongoing and upcoming high-resolution surveys are expected to significantly enhance our understanding of the progenitors and explosion mechanisms of SNe.
Looking ahead, the Vera Rubin Observatory’s Legacy Survey of Space and Time (LSST) is expected to detect faint sources down to $r \sim 27.5\,$mag and discover up to 10 million SNe over its nominal 10-year mission \citep{2025A&A...699A..98S}.
Meanwhile, the China Space Station Telescope (CSST), a serviceable two-meter-aperture wide-field telescope, is expected to provide over 16000 well-classified SN candidates, and its near-ultraviolet observations of CSST are anticipated to capture hundreds of shock-cooling events each year \citep{2024SCPMA..6719512L}.
With the increasing availability of deep, high-cadence survey data, these large-scale telescope observations will provide critical insights into the diverse origins and evolutionary pathways of SNe.

\section*{Data availability}

Photometric data for this SN IIb presented in this study are available at the CDS \url{https://cdsarc.cds.unistra.fr/viz-bin/cat/J/A+A/704/A233}.
Our spectra observations are available via the Weizmann Interactive Supernova Data Repository \citep[WISeREP; ][]{2012PASP..124..668Y} at \url{https://www.wiserep.org/object/2412}.

\bibliographystyle{aa} 
\bibliography{aa56873-25.bib} 
--------------------------------------------------------
\FloatBarrier
\clearpage

\appendix
\onecolumn

\section{Supplementary figures}\label{sect:supfigure}

\begin{figure}[H]
   \centering
   \includegraphics[width = 0.6\textwidth]{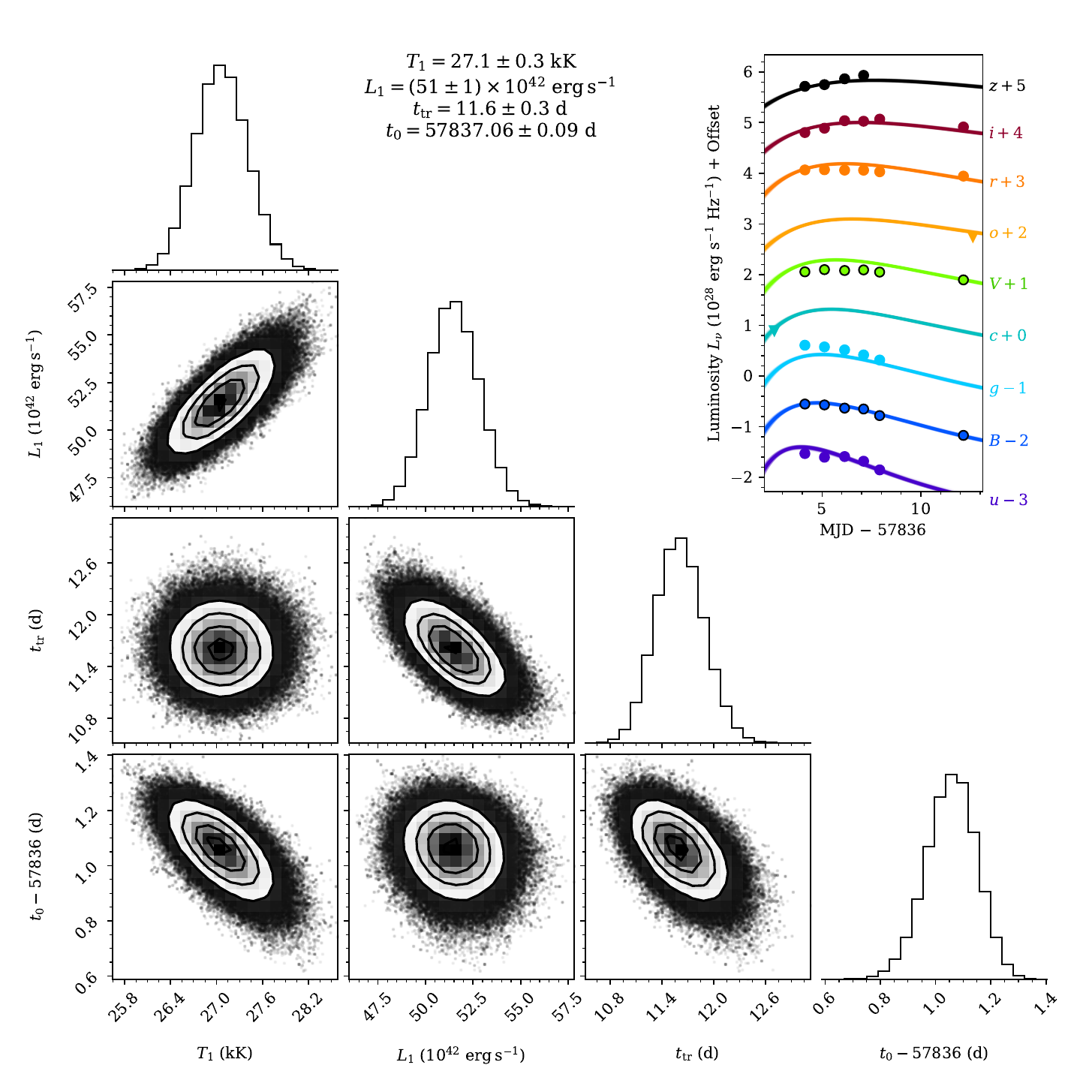}
   \caption{
  Modelling of shock-cooling light curve of SN\,2017ckj using Light Curve Fitting package \citep{hosseinzadeh_2024_11405219} based on the shock-cooling model of \citet{2017ApJ...838..130S}.
   The posterior probability distributions of the model parameters are shown, including the temperature 1 day after explosion ($T_1$), the total luminosity 1 day after explosion ($L_1$), the time at which the envelope becomes transparent ($t_{tr}$), and the time of explosion ($t_0$).
    The best-fit values of these parameters and their 1$\sigma$ credible intervals are provided at the top.
   }
    \label{Fig:shockcooling_fitting.pdf}
\end{figure} 

\begin{figure}[H]
    \centering
    \begin{subfigure}{0.5\textwidth}
        \centering
        \includegraphics[width=\linewidth]{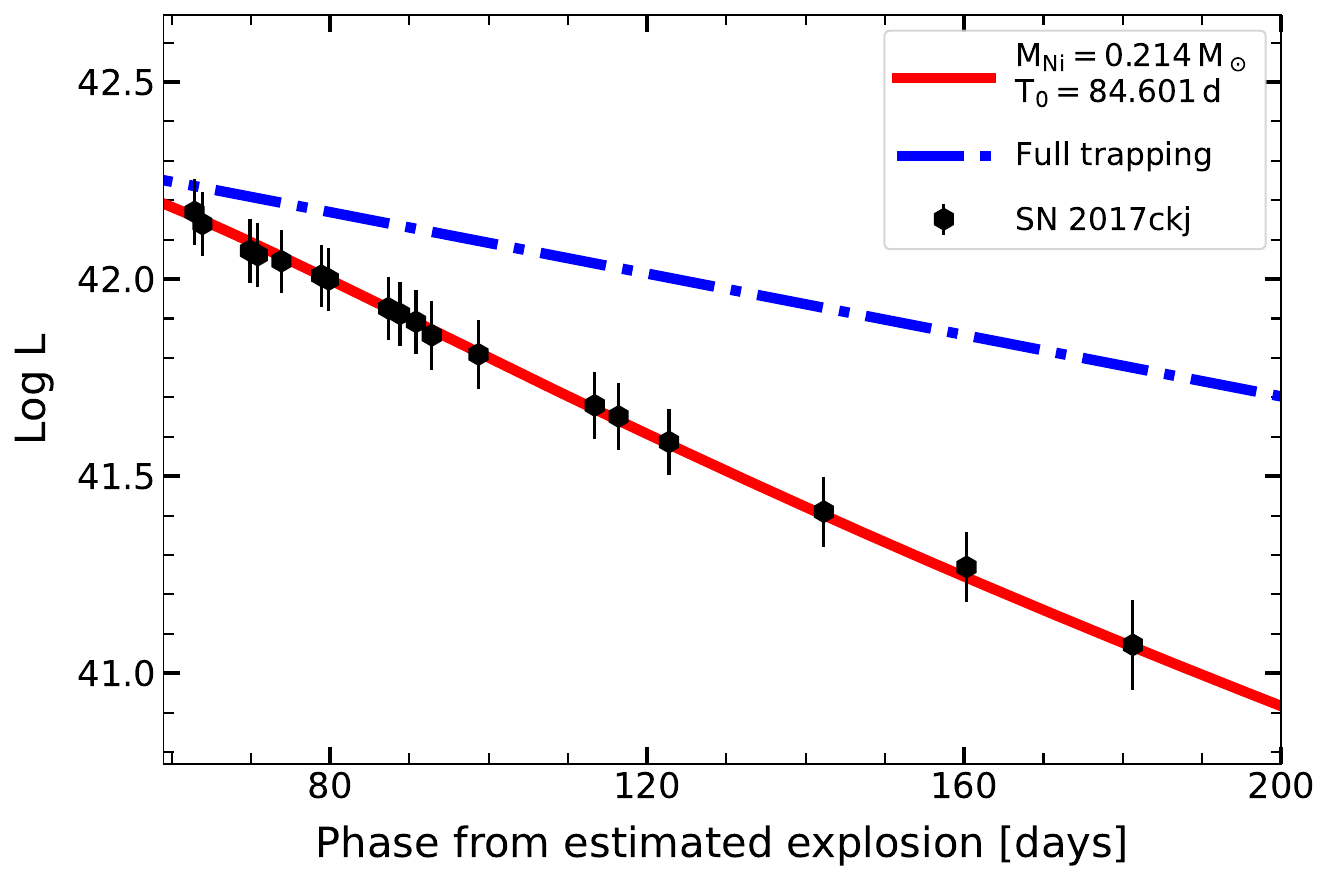}
        \label{fig:left}
    \end{subfigure}
    \begin{subfigure}{0.35\textwidth}
        \centering
        \includegraphics[width=\linewidth]{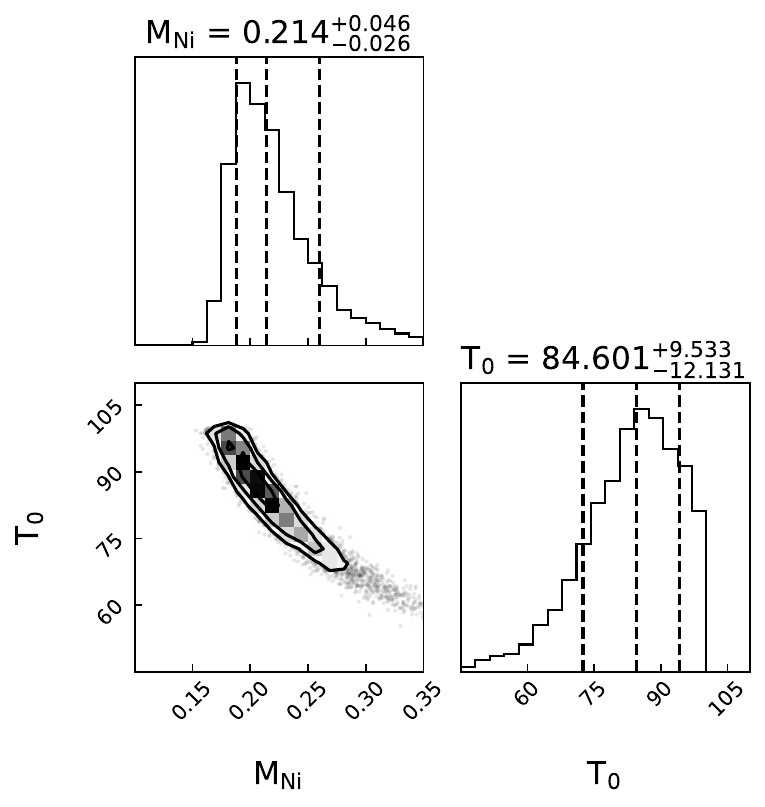}
        \label{fig:right}
    \end{subfigure}
    \caption{Fit of the modified radioactive decay model to the late-time full bolometric light curve of SN\,2017ckj. 
    In the left panel, the bolometric light curve fitting is shown by the red solid line, while the full $\gamma$-ray trapping model is presented in blue dashed-dotted line for comparison.
    The right panel presents the posterior distribution from MCMC sampling.
    The mean of the posterior distribution and 1$\rm \sigma$ uncertainties are marked.}
    \label{fig:latebolo fitting}
\end{figure}

\clearpage
\twocolumn

\section{Observations and data reduction} \label{sect:data reduction}

\subsection{Photometric data}
We conducted multi-band optical (Sloan $ugriz$, Johnson-Cousins $BV$) follow-up campaigns of SN\,2017ckj starting shortly after its classification. 
The telescopes and instruments utilised were the following:
The 1.82-m Copernico Telescope atop Mount Ekar with the Asiago Faint Object Spectrograph and Camera (AFOSC), hosted by INAF -- Padova Astronomical Observatory, at the Asiago site, Italy;
The 2.56-m Nordic Optical Telescope (NOT) equipped with the Alhambra Faint Object Spectrograph and Camera (ALFOSC), and the 10.4-m Gran Telescopio Canarias (GTC) with Optical System for Imaging and low-Intermediate-Resolution Integrated Spectroscopy (OSIRIS), both located at Roque de los Muchachos Observatory (La Palma, Canary Islands, Spain).

The raw images were first pre-reduced by applying bias and overscan corrections, flat-fielding, and trimming, which are standard correction steps performed in \textsc{iraf}\footnote{\url{https://iraf-community.github.io/}} \citep{Tody1986SPIE..627..733T, Tody1993ASPC...52..173T}. 
If multiple exposures were taken with the same instrument and on the same night, we combined them into stacked science frames to increase the S/N.
The steps necessary to obtain the SN magnitude were performed using the dedicated pipeline {\sc snoopy}\footnote{\url{https://sngroup.oapd.inaf.it/ecsnoopy.html}} developed by \citet{snoopyref}.
{\sc snoopy} consists of a collection of PYTHON-based scripts calling standard {\sc iraf} tasks within {\sc pyraf}.
The template subtraction was adopted for $ugriz$ bands and $BV$ bands of SN\,2017ckj to remove the background contamination from the host galaxy. 
When the SN was not detected, an upper limit to the object brightness was estimated. 
Photometric calibration of instrumental magnitudes was performed by adopting instrumental zero points (ZPs) and colour terms (CTs) inferred through observations of standard stars on photometric nights.
Specifically, Johnson-Cousins filter photometry was calibrated using standard stars from the \citet{Landolt1992AJ....104..340L} catalogue, while Sloan-filter data were retrieved from the SDSS DR 18 catalogue \citep{Abdurro'uf2022ApJS..259...35A}. 
To correct the instrumental ZPs on non-photometric nights and improve the photometric calibration accuracy, we compared the average magnitudes of local sequences of standard stars in the fields of the SN to those obtained on photometric nights.
With the corrected ZPs, we calibrated the SN apparent magnitudes on all nights. 


We also collected photometric data from the public ATLAS sky surveys for transients.
The orange ($o$) and cyan ($c$) band light curves were directly produced by the ATLAS data-release server \footnote{\url{https://fallingstar-data.com/forcedphot/}} \citep{Shingles2021TNSAN...7....1S}. 
Panoramic Survey Telescope and Rapid Response System (Pan-STARRS) is a wide-field imaging survey system \citep{2016arXiv161205560C} and also provides late-time photometric data, which we incorporate in this paper.

\subsection{Spectroscopic data}

Spectroscopic observations of the SN\,2017ckj were carried out using the following telescopes:
Copernico 1.82-m/AFOSC, 2.56-m NOT/ALFOSC, and 10.4-m GTC/OSIRIS.
The spectra obtained with Copernico 1.82-m/AFOSC, 2.56-m NOT/ALFOSC, and GTC/OSIRIS were reduced using the dedicated pipeline \textsc{Foscgui} developed by E. Cappellaro.
All raw spectral data were reduced following the standard steps in \textsc{iraf}\footnote{\url{https://iraf.readthedocs.io/en/latest/}} \citep{Tody1986SPIE..627..733T, Tody1993ASPC...52..173T}.
The pre-reduction steps, such as bias, overscan, flat-fielding correction, and trimming, are similar to those described for the imaging data.
Then, the one-dimensional (1D) spectra were optimally extracted from the 2D images. Wavelength calibrations were performed using arc lamps, while flux calibrations were performed using spectrophotometric standard stars taken on the same nights.
Subsequently, the strongest telluric absorption bands, such as O$_2$ and H$_2$O, were removed from the SN spectra using the spectra of the standard stars.
Finally, the accuracy of flux calibration for all spectra was checked against the coeval photometric data.
The information on the instrumentation used for the spectroscopic observations is reported in Table~\ref{Tab:17ckjspecinfo}.

\section{Data tables} \label{Sect:datainfo}

\begin{table*}[t]
    \centering
    \small
    \setlength{\tabcolsep}{6pt}  
    \caption{Properties of the comparison SNe IIb.}
    \begin{tabular}{ccccccc}
        \hline
        SN IIb & Explosion Date & Redshift & Distance & $E(B-V)_\mathrm{Gal}$ & $E(B-V)_\mathrm{Host}$ & References \\
         & [MJD] & z & [Mpc] & [mag] & [mag] & \\
        \hline
        1993J & 49072.0 & -0.000113 & 2.9 & 0.069 & 0.11 & 1,2,3 \\
        2008ax & 54528.8 & 0.00456 & 9.6$\pm$1.3 & 0.022 & 0.278 & 4,5,6 \\
        2011dh & 55712.5 & 0.001638 & 8.03$\pm$0.77 & 0.03 & 0.04 & 7,8,9 \\
        2011fu & 55824.5 & 0.001845 & 74.5$\pm$5.2 & 0.068 & 0.035 & 10 \\
        2013cu & 56414.93 & 0.025734 & 108 & 0.0105 & 0 & 11 \\
        DES14X2fna & 56927.7 & 0.0453 & 200.7$\pm$3.0 & 0.0225 & 0 & 12\\
        2015as & 57332.0 & 0.0036 & 19.2 & 0.008 & 0 & 13 \\ 
        2016gkg & 57651.2 & 0.0049 & 21.8 & 0.0166 & 0.09 & 14,15 \\
        2018gk / ASASSN-18am & 58130.1 & 0.031010 & 140.5$\pm$2.3 & 0.0086 & 0 & 16\\
        2019tua & 58785.3 & 0.010 & 41.39 & 0.06 & 0 & 17\\
        2020acat & 59192.0 & 0.007932 & 35.3$\pm$4.4 & 0.0207 & 0 & 18 \\
        2021bxu & 59246.3 & 0.0178 & 72$\pm5$ & 0.014 & 0 & 19\\
        2017ckj & 57836.6 & 0.037 & 158.1$\pm11.1$ & 0.013 & 0 & This work \\
        \hline
    \end{tabular}%
    \begin{flushleft}
    \small
    References: 
    1. \citet{1994AJ....107.1022R},
    2. \citet{1995A&AS..110..513B},
    3. \citet{1996AJ....112..732R},
    4. \citet{2008MNRAS.389..955P},
    5. \citet{2009PZ.....29....2T},
    6.\citet{2011MNRAS.413.2140T},
    7. \citet{tsvetkov2012photometric},
    8. \citet{2013MNRAS.433....2S}, 
    9. \citet{2014A&A...562A..17E}, 
    10. \citet{2015MNRAS.454...95M}, 
    11. \citet{2014Natur.509..471G},
    12. \citet{2021MNRAS.505.3950G},
    13. \citet{2018MNRAS.476.3611G},
    14. \citet{2017ApJ...837L...2A}, 
    15. \citet{2018Natur.554..497B},
    16. \citet{2021MNRAS.503.3472B},
    17. \citet{2024ApJ...970..103H},
    18. \citet{2022MNRAS.513.5540M},
    19. \citet{2023MNRAS.524..767D}.
    \end{flushleft}
    \label{tab:SNe_IIb info}
\end{table*}

\begin{table*}[t]
\caption{Log of spectroscopic observations of SN\,2017ckj.}
\label{Tab:17ckjspecinfo}
\centering
\small                                      
\setlength{\tabcolsep}{6pt}
\begin{tabular}{c c c c c c c c} 
\hline
Date & MJD & Phase$^a$ & Telescope+Instrument & Grism/Grating+Slit & Spectral range & Resolution & Exp. time \\ 
  &   & (days) &   &        & (\AA)    & (\AA)           & (s)           \\ 
\hline 
20170328 & 57840.1  & +3.0  & Copernico+AFOSC & gr04+1.69" & 2890-7950  & 16 & 1800        \\
20170329 & 57841.1  & +4.0  & Copernico+AFOSC & vph7+1.69" & 3230-8950  & 16 & 1800       \\
20170330 & 57842.2  & +5.1  & Copernico+AFOSC & vph7+1.69"  & 3470-7030 & 16 & 1800       \\
20170331 & 57843.1  & +6.0  & Copernico+AFOSC & vph7+1.69" & 3290-7030  & 16  & 1800        \\
20170405 & 57848.1  & +11.0 & NOT+ALFOSC   & gm4+1.0"  & 3460-9310  & 12 & 1800      \\
20170422 & 57865.2  & +28.1 & NOT+ALFOSC   & gm4+1.3"  & 3660-9310  & 14 & 2400      \\
20170506 & 57879.0  & +41.9 & NOT+ALFOSC   & gm4+1.0"  & 3660-9300  & 12 & 3600      \\
20170531 & 57904.1  & +67.0 & NOT+ALFOSC   & gm4+1.0"  & 3760-9410  & 12 & 3600      \\
20170627 & 57931.0  & +93.9 & NOT+ALFOSC   & gm4+1.0"  & 3660-9360  & 12 & 2700      \\
20170715 & 57950.0 & +112.9 & GTC+OSIRIS & R1000B/R1000R+1.0" & 3750-10020 &8 & 1800 \\
20170810 & 57976.0 & +138.9 & GTC+OSIRIS & R300B/R300R+1.0" & 3880-10260 &7 & 1800 \\
\hline
\multicolumn{7}{l}{{$^a$Phases are relative to the estimated explosion epoch (MJD = $57837.1\, \pm 0.1$) in observer frame.}} \\
\end{tabular}
\end{table*}

\section*{Acknowledgements}
We gratefully thank the anonymous referee for his/her insightful comments and suggestions that improved the paper.

This work is supported by the National Natural Science Foundation of China (Nos 12288102, 12225304, 12090040/12090043, 12303054, 12473032), the National Key Research and Development Program of China (Nos. 2021YFA1600404, 2024YFA1611603), the Western Light Project of CAS (No. XBZG-ZDSYS-202117), the Yunnan Revitalization Talent Support Program (Yunling Scholar Project), the Yunnan Revitalization Talent Support Program—Young Talent project, the Yunnan Fundamental Research Projects (Nos 202201BC070003, 202401AU070063, 202501AS070078, 202501AW070001, 202501AS070005), and the International Centre of Supernovae, Yunnan Key Laboratory (No. 202302AN360001).
SB, EC, NER, PO, AP, AR, IS, LT and GV acknowledge financial support from the PRIN-INAF 2022 "Shedding light on the nature of gap transients: from the observations to the models". AR also acknowledges financial support from the GRAWITA Large Program Grant (PI P. D’Avanzo).
AF acknowledges funding by the European Union – NextGenerationEU RFF M4C2 1.1 PRIN 2022 project ``2022RJLWHN URKA'' and by INAF 2023 Theory Grant ObFu 1.05.23.06.06 ``Understanding R-process \& Kilonovae Aspects (URKA)''.
JH acknowledges the Vilho, Yrjö and Kalle Väisälä Foundation of the Finnish Academy of Science and Letters.
TK acknowledges support from the Research Council of Finland project 360274.
T.M.R is part of the Cosmic Dawn Center (DAWN), which is funded by the Danish National Research Foundation under grant DNRF140. T.M.R acknowledges support from the Research Council of Finland project 350458.
S.M. acknowledges financial support from the Research Council of Finland project 350458.
We thank S. Taubenberger for conducting the template observations with the Copernico Telescope.
We thank Jingxiao Luo, Xinbo Huang and Zeyi Zhao for helpful discussions and valuable suggestions.

Based on observations made with the Gran Telescopio Canarias (GTC; Program GTC-17A), installed at the Spanish Observatorio del Roque de los Muchachos of the Instituto de Astrofísica de Canarias, on the island of La Palma.
Based on observations collected with the AFOSC instrument at the 1.82-m Copernico Telescope of the INAF – Osservatorio Astronomico di Padova (Asiago, Italy).
We thank the technical staff of the Asiago Observatory for their support during the observations.
The data in this study include observations made with the Nordic Optical Telescope, owned in collaboration by the University of Turku and Aarhus University and operated jointly by Aarhus University, the University of Turku, and the University of Oslo, representing Denmark, Finland, and Norway; the University of Iceland; and Stockholm University at the Observatorio del Roque de los Muchachos, La Palma, Spain, of the Instituto de Astrofisica de Canarias.
The data presented here were obtained in part with ALFOSC, which is provided by the Instituto de Astrofísica de Andalucía (IAA) under a joint agreement with the University of Copenhagen and NOT.
This work has made use of data from the Asteroid Terrestrial-impact Last Alert System (ATLAS) project. The ATLAS project is primarily funded to search for near earth asteroids through NASA grants NN12AR55G, 80NSSC18K0284, and 80NSSC18K1575; byproducts of the NEO search include images and catalogs from the survey area.
The ATLAS science products have been made possible through the contributions of the University of Hawaii Institute for Astronomy, the Queen’s University Belfast, the Space Telescope Science Institute, the South African Astronomical Observatory, and The Millennium Institute of Astrophysics (MAS), Chile.
This work has made use of data from the Pan-STARRS1 Surveys (PS1) and the PS1 public science archive. 
The PS1 and the PS1 public science archive have been made possible through contributions by the Institute for Astronomy, the University of Hawaii, the Pan-STARRS Project Office, the Max-Planck Society and its participating institutes, the Max Planck Institute for Astronomy, Heidelberg and the Max Planck Institute for Extraterrestrial Physics, Garching, The Johns Hopkins University, Durham University, the University of Edinburgh, the Queen's University Belfast, the Harvard-Smithsonian Center for Astrophysics, the Las Cumbres Observatory Global Telescope Network Incorporated, the National Central University of Taiwan, the Space Telescope Science Institute, the National Aeronautics and Space Administration under Grant No. NNX08AR22G issued through the Planetary Science Division of the NASA Science Mission Directorate, the National Science Foundation Grant No. AST–1238877, the University of Maryland, Eotvos Lorand University (ELTE), the Los Alamos National Laboratory, and the Gordon and Betty Moore Foundation.
This research has made use of the NASA/IPAC Extragalactic Database (NED), which is operated by the Jet Propulsion Laboratory, California Institute of Technology, under contract with the National Aeronautics and Space Administration. iraf was distributed by the National Optical Astronomy Observatory, which was managed by the Association of Universities for Research in Astronomy (AURA), Inc., under a cooperative agreement with the U.S. NSF.

\end{document}